\documentclass[%
 reprint,superscriptaddress,
 amsmath,amssymb,
 aps
]{revtex4-2}
\usepackage{graphicx}
\usepackage{dcolumn}
\usepackage{bm}
\usepackage{xcolor}
\usepackage{wick}
\usepackage{graphicx}
\usepackage{amsmath}
\usepackage{amssymb}
\usepackage{soul}
\usepackage[version=4]{mhchem}
\usepackage{siunitx}
\usepackage{longtable,tabularx}
\usepackage{braket}
\usepackage{wick}
\usepackage{simplewick}
\begin{document}
\preprint{APS/123-QED}
\title{Accurate Determination of Relativistic Effects in Atoms by Quantum Annealing}
\title{Accurate Computation of Relativistic Excitation Energies Using Quantum Annealing}
\author{Vikrant Kumar}
\email{vikrantkumar.895@gmail.com}
\affiliation{Centre for Quantum Engineering, Research and Education, TCG CREST, Salt Lake, Kolkata 700091, India}
\author{Nishanth Baskaran}
\affiliation{Centre for Quantum Engineering, Research and Education, TCG CREST, Salt Lake, Kolkata 700091, India}
\author{V. S. Prasannaa}
\affiliation{Centre for Quantum Engineering, Research and Education, TCG CREST, Salt Lake, Kolkata 700091, India}
\author{K. Sugisaki}
\affiliation{Centre for Quantum Engineering, Research and Education, TCG CREST, Salt Lake, Kolkata 700091, India}
\affiliation{Department of Chemistry, Graduate School of Science, Osaka Metropolitan University, 3-3-138 Sugimoto, Sumiyoshi-ku, Osaka 558-8585, Japan}
\affiliation{JST PRESTO, 4-1-8 Honcho, Kawaguchi, Saitama 332-0012, Japan}
\author{D. Mukherjee}
\affiliation{Centre for Quantum Engineering, Research and Education, TCG CREST, Salt Lake, Kolkata 700091, India}
\author{K.G. Dyall}
\affiliation{Dirac Solutions, Portland, Oregon 97229, USA}
\author{B. P. Das}
\affiliation{Centre for Quantum Engineering, Research and Education, TCG CREST, Salt Lake, Kolkata 700091, India}
\affiliation{Department of Physics, Tokyo Institute of Technology, 2-12-1 Ookayama, Meguro-ku, Japan}

\begin{abstract}

 We report the first results for the computation of relativistic effects in quantum many-body systems using quantum annealers. An average accuracy of 98.9\% in the fine structure splitting of boron-like ions with respect to experiments has been achieved using the Quantum Annealer Eigensolver~(QAE) algorithm on the D-Wave Advantage hardware. We obtain these results in the framework of the many-electron Dirac theory. We implement QAE using a hybrid quantum annealing method that includes a novel qubit encoding scheme and decomposing the problem into smaller ones based on perturbation theory.
\end{abstract}

\maketitle
\section{Introduction}

Quantum annealing~(QA) is a metaheuristic method for solving optimization problems using quantum effects~\cite{Hauke2020PerspectivesImplementations}. It was formulated in its current form by Kadowaki and Nishimori~\cite{NishimoriQA1998} and is steadily gaining in importance. It is related to Adiabatic Quantum Computing, which is considered to be the second paradigm of quantum computing~\cite{Yarkoni2022QuantumReview, Chakrabartireview2023, Crosson2021review, Canivell2021qilimanjaro, APSmeetingweber}. Applications of QA include solving a range of problems ~\cite{Mott2017SolvingLearning, Li2018QuantumProblem, Irie2021HybridDynamics, KingCoherentChain, Kitai2020DesigningMachines, Inoue2021TrafficAnnealing, Matsuura_2020}, including the electronic structure of atoms and molecules~\cite{Xia2018ElectronicHamiltonian, Streif2019SolvingAnnealer, Teplukhin2019CalculationAnnealer, Teplukhin2020ElectronicAnnealer, Teplukhin2021ComputingAnnealer}. The recently proposed Quantum Annealer Eigensolver (QAE) algorithm is a promising direction for computing molecular vibrational spectra~\cite{Teplukhin2019CalculationAnnealer}, ground and excited state energies~\cite{Teplukhin2020ElectronicAnnealer,Teplukhin2021ComputingAnnealer}. QAE solves an eigenvalue problem by minimizing a suitable objective function by using the Rayleigh-Ritz variational principle, and has potential applications in many areas of science and engineering and beyond~\cite{Eigengenome, Eigendiffusive, eigeninfection, eigenfluids, eigenfaces}.

Relativistic effects play a crucial role in various physical and chemical phenomena~\cite{Pyykko2012RelativisticThought,AhujaRelativity2011}. Their inclusion in quantum many-body calculations is a major challenge, and has applications in many atomic and molecular problems concerning new physics beyond the standard model of elementary particles~\cite{Dzuba2022RelativisticConstant, Flambaum2019OscillatingRotation, Sahoo2016EnhancedMoment, PhysRevLett.120.203001, Sahoolorentzinvariance2019, Yuopticalclocks}. 

The QAE algorithm can determine the fine structure splitting~(FSS) by taking the difference between the minimum energies of two sets of atomic states with the same orbital angular momenta~(L, S) but different total angular momentum, J. In boron and its isoelectronic systems, the $^2P$ ground state splits into two odd parity states, one with $J=\frac{1}{2}$ and the other with $J=\frac{3}{2}$. In this letter, we estimate the lowest energy state of each set using QAE~(simulation and hardware), and the corresponding difference allows us to evaluate the FSS in boron-like ions, which we compare with the experiment. This is the first time that QA has been used to accurately calculate a purely relativistic property. 

\begin{figure*}[t]
\centering
\includegraphics[scale = 0.11]{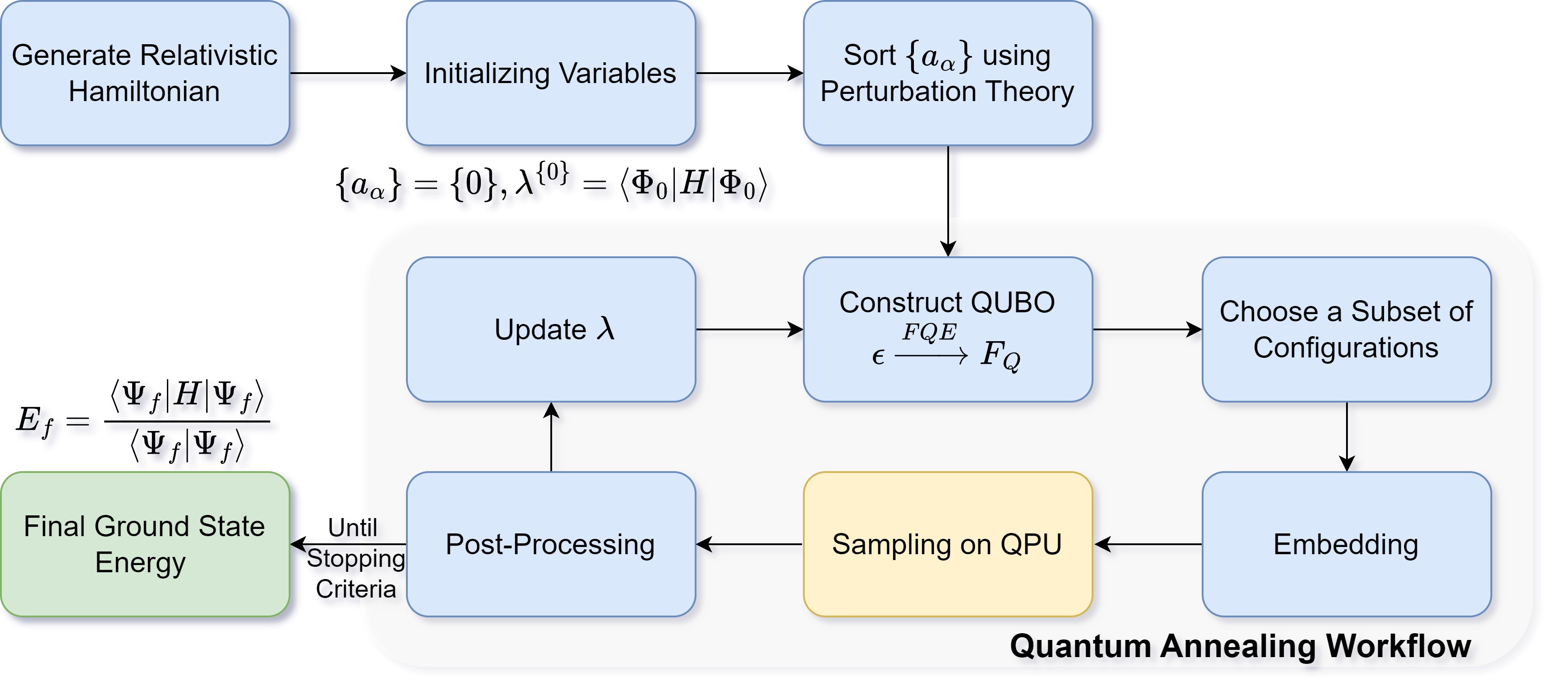} 
\caption{\label{fig:FIG1} Illustration of our QAE workflow. The Lagrange multiplier, $\lambda^{\{0\}}$, is initialized to the CSF energy associated with the most dominant configuration, $\langle \Phi_0 | H| \Phi_0 \rangle$. $\{ a_\alpha\}$ denotes the list of expansion coefficients from the relCI expansion of the wave function. $E_f$ and $\ket{\Psi_{f}}$ refer to the final ground-state energy and its corresponding wavefunction obtained using the presented workflow, respectively. Floating qubit encoding is abbreviated as FQE in the figure. }
\end{figure*}

FSS is relativistic in origin and is influenced by electron correlation effects~\cite{Das1984Ground-stateSequence, Das2011RELATIVISTICATOMS}, which are challenging to capture accurately because of the lack of all-to-all connectivity between the qubits of current quantum annealers. Recently, hybrid workflows with quantum annealing and classical components such as qbsolv~\cite{qbsolv} have partially alleviated this drawback through decomposition methods, which entail dividing the problem into smaller parts. We adapt the QAE algorithm with certain key workflow improvements to compute the FSS as depicted in Fig.~\ref{fig:FIG1}. Given that the FSS arises entirely due to relativity, we choose moderately heavy boron-like ions in this work, for which the interplay of relativistic and correlation effects is crucial. Experimental data is available for the specific ions that we have chosen for our computations~\cite{Edlen1983ComparisonSequence, Hinnov1980MagneticNickel, PhysRevA.40.1488, Myrnas1994TransitionsTokamak}. 

\section{Theory and workflow}

The dominant relativistic effects in atomic systems are contained in the Dirac-Coulomb Hamiltonian~\cite{2007DC}, given as 
\begin{equation}
    \label{HDC}
    H_{DC} =\sum_i\left(c \alpha_i \cdot p_i + \beta_i c^2 + V_N(r_i)\right) + \sum_{j>i}\frac{1}{r_{ij}}.
\end{equation}
The lowest-order relativistic correction to the Coulomb interaction is known as the Breit interaction~\cite{2007B} and whose Hamiltonian is
\begin{equation}
    \label{HB}
    H_B = - \sum_{j>i} \left( \frac{\alpha_i \cdot \alpha_j}{2r_{ij}} + \frac{(\alpha_i \cdot r_{ij})(\alpha_i \cdot r_{ij})}{2r_{ij}^3} \right),
\end{equation}
where both $H_{DC}$ and $H_{B}$ are given in atomic units, $V_N(r_i)$ is the potential due to the nucleus, $\alpha$ and $\beta$ are the Dirac matrices, and $c$ is the velocity of light. $r_{ij}$ is the inter-electronic separation. The Dirac-Coulomb-Breit~(DCB) Hamiltonian is the sum of the above two Hamiltonians.  

The first step in the QAE algorithm is the determination of the matrix elements of the DCB Hamiltonian on a traditional computer. For this purpose, we employ a well established relativistic atomic structure code ~\cite{Parpia1991SoftwareOxford}. The matrix elements are computed using suitable configuration state functions~(CSFs) corresponding to a specific angular momentum (in our case, either $J=\frac{1}{2}$ or $\frac{3}{2}$) and parity (odd, in this work) as basis functions. These  CSFs are built by considering the complete active space (CAS) consisting of single-particle ($1s$, $2s$, $2p_{1/2}$ and $2p_{3/2}$) orbitals. The orbitals are evaluated using a state-averaged calculation within the multi-configuration Dirac-Fock~(MCDF) method ~\cite{Das1984Ground-stateSequence}, where we have used a \textit{common} set of optimized orbitals for the $J= \frac{1}{2}$ and $J=\frac{3}{2}$ cases. We extract two sub-matrices from the CAS Hamiltonian matrix, one built from odd-parity CSFs with $J= \frac{1}{2}$, while the other is from $J=\frac{3}{2}$ CSFs, thereby obtaining matrices of size $(9 \times 9)$ and $(16 \times 16)$, for the $^2P_{1/2}$ and the $^2P_{3/2}$ states respectively. 

The energy functional ($\epsilon$) of interest to us, is given by 
\begin{equation}
    \label{ep1}
    \epsilon = \langle \Psi |H_{DCB}|\Psi\rangle - \lambda \langle \Psi |\Psi\rangle, 
\end{equation}
where $\lambda$ refers to the Lagrange multiplier that guarantees the normalization of the wave function. The wave function can be expressed as a linear combination of CSFs, $|\Psi \rangle = \sum_{\alpha=0}^{\mathcal{B}-1} a_{\alpha} | \Phi_{\alpha} \rangle$, where the $a_{\alpha}$'s denote the expansion coefficients corresponding to the CSFs. Upon minimizing the energy functional with respect to the expansion coefficients, we can obtain the ground state energy. The procedure for optimizing $\lambda$ is discussed in the Section~\ref{supp_sec1} of the Supplementary Material. Substituting $|\Psi \rangle$ in Eq~(\ref{ep1}), we obtain 

\begin{equation}
    \epsilon = \sum_{\alpha,\beta}^{\mathcal{B}-1}a_\alpha a_\beta H_{\alpha \beta} - \lambda \sum_{\alpha}^{\mathcal{B}-1}a_\alpha^2,
\end{equation}
where $H_{\alpha \beta} = \langle \Phi_\alpha|H| \Phi_\beta \rangle$. To convert the expression to its quadratic unconstrained binary optimization (QUBO) form for making it compliant with the D-Wave hardware, we represent $a_\alpha$ in the $i^{th}$ iteration, denoted by $a_{\alpha}^{\{i\}}$, in binary using $K$ bits through our qubit encoding scheme given by

\begin{eqnarray}
\label{fFPR}
        a_\alpha^{\{i\}} &=& \mu_\alpha + \sigma \sum_{k=0}^{K-1}f_k 2^{-k}q_k^\alpha; 
\end{eqnarray}

 where $f_k=-1$ if $k=0$ and 1 otherwise, and $q_i$ are binary variables. $\mu_\alpha$ and $\sigma$ are updated in each iteration 
\begin{eqnarray}
        \mu_\alpha &=& a_\alpha^{\{i-1\}}\, \textrm{and} \nonumber \\
        \sigma &=& 2^{(-i+1)/2}, \quad \forall \alpha. \nonumber
\end{eqnarray}

The initial guess for ${a_\alpha}$ is set to zero. The second term in Eq~\eqref {fFPR} is used as a correction to the previous estimate of $a_\alpha$ (stored as $\mu_\alpha$, which is shifted appropriately with $\sigma$ to ensure that ${a_\alpha}$ is within $[-1,1]$). This is because $\sigma$ decreases by a factor of $\frac{1}{\sqrt{2}}$ with each iteration, allowing for increasingly smaller values of $a_\alpha$ to be searched for the minimum. The qubit encoding scheme adopted in the present work requires fewer qubits than previous works~\cite{Teplukhin2019CalculationAnnealer}, but the optimization part is performed in an iterative manner till a precision of $10^{-5}$ is achieved. Fixed point encoding (as used in previous QAE approaches~\cite{Teplukhin2019CalculationAnnealer}) limits the precision of each $a_\alpha$ by the number of qubits used to represent them, and increasing the number of qubits beyond 10 does not improve results further due to errors associated with chaining. Hence, capturing coefficients below $10^{-3}$ becomes difficult with fixed point encoding. 

The mapping scheme discussed above increases the Hamiltonian size from $(\mathcal{B} \times \mathcal{B})$ to $(\mathcal{B}K \times \mathcal{B}K)$. The explicit QUBO form after the above conversion (refer to Section~\ref{secII_qubo_deri} of Supplementary Material) is given as

\begin{equation}
\label{FQ}
    F_Q= \sum_{\alpha, \beta = 0}^{\mathcal{B}-1}\left( \sum_{n,m = 0}^{K-1}q_n^\alpha A q_m^\beta + \sum_{n = 0}^{K-1} B q_n^\alpha + C\right),
\end{equation}

\noindent where
\begin{eqnarray}
A &=& (\sigma^2 f_n f_m 2^{-(n+m)})H_{\alpha \beta}', \nonumber \\ 
B &=& (2\mu_\beta \sigma f_n 2^{-n})H_{\alpha \beta}', \nonumber \\ 
C &=& \mu_\alpha \mu_\beta H_{\alpha \beta}', 
\text{ and } \nonumber \\ 
H_{\alpha \beta}' &=& (H_{\alpha \beta}-\lambda   \delta_{\alpha \beta}). \nonumber
\end{eqnarray}
\noindent In the above equation, $\delta_{nm}$ is 1 if $n=m$ and 0 otherwise. $A$ and $B$ map to the quadratic biases~(couplings) and linear biases of the qubits, respectively.
The above QUBO is usually large in size and densely connected, making it difficult to solve on current D-Wave hardware. To address this issue, we employ hybrid quantum annealing to perform optimizations of sub-problems, i.e. Eq.~\eqref{FQ} corresponding to a subset of coefficients, for a particular $\lambda$ iteratively. Each such iteration will henceforth be mentioned as a `Repeat'. A Repeat therefore involves decomposition, embedding, annealing, composition and post-processing (in the order mentioned). Decomposition reduces the size of the problem by selecting a subset of coefficients corresponding to configurations that arise from the first-order perturbation of the Dirac-Fock state. Performing optimization in a reduced space allows the annealer to yield results that are more accurate than optimizing all the coefficients together. Prior to quantum annealing, the sub-problem is embedded to the D-Wave annealer's architecture. $F_Q$ is then optimized through repeated anneals to output qubit configurations that are close to the ground state of $F_Q$. The qubit samples thus obtained from the annealer are then merged with the optimal qubit configuration (initially a list of zeros), then scaled accordingly, and finally compared against the previous best solution. A more detailed description of the steps can be found in Section~\ref{supp_sec1} of the Supplementary Material. In comparison to the default decomposition strategy in D-Wave hybrid software, where the priority of each qubit is based on its effect on the QUBO energy $F_Q$, our strategy obtains better results in significantly fewer Repeats (refer to Table~\ref{S_ED} in Supplementary Material), and also has lower classical overhead since the priority of each configuration is computed once and is used throughout the QAE computation, whereas the default method computes the priority for each qubit for every Repeat. We carried out our implementation using the existing modules of D-Wave’s Ocean~\cite{dwaveoceans} and hybrid~\cite{dwavehybrid} software development kits. Throughout this work, $K=10$. The workflow parameters are presented in Table~\ref{table-params_hardware} of the Supplementary Material. 

\begin{figure}
    \centering
    \includegraphics[scale=0.75]{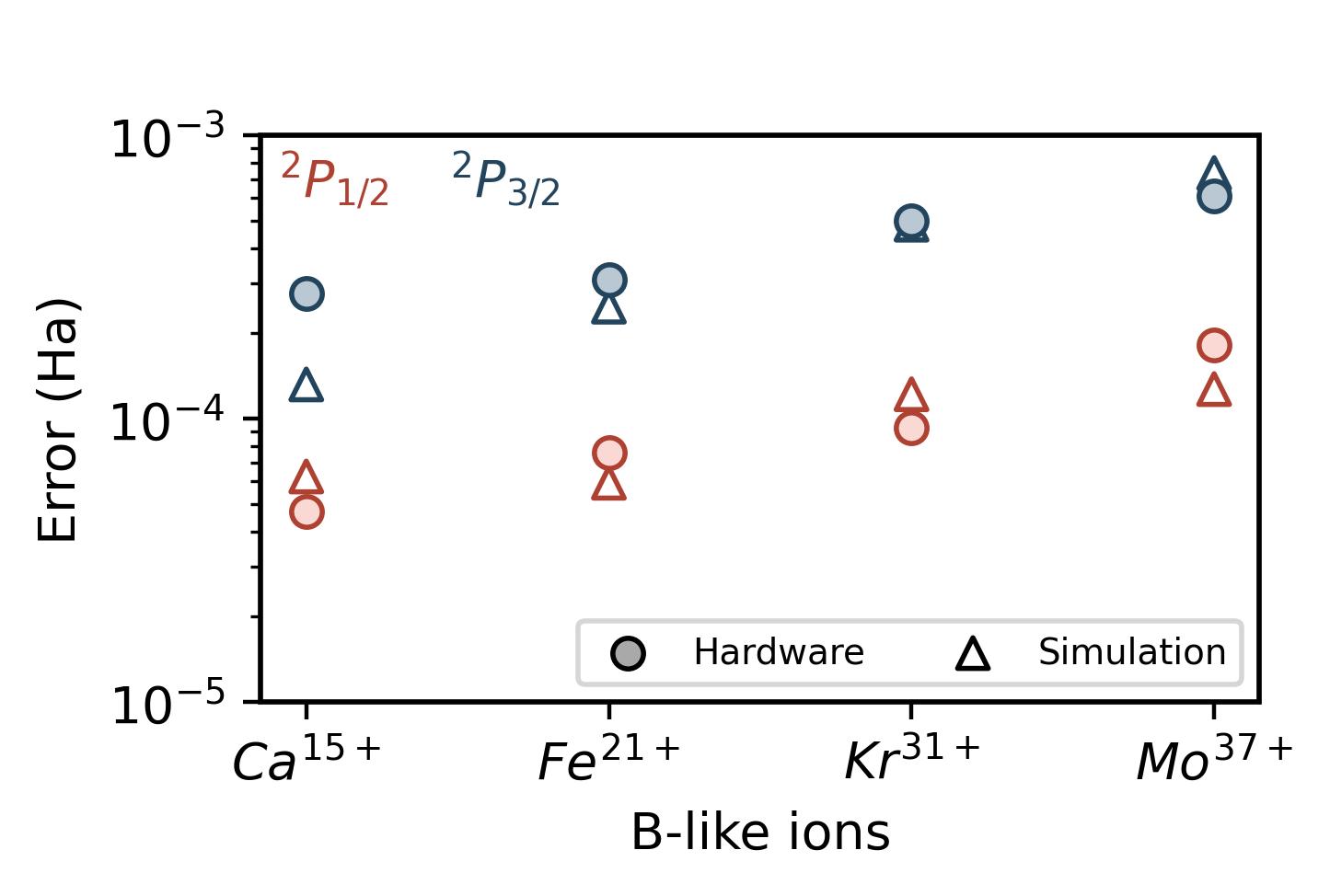}
    \caption{Error with respect to relCI in QAE with Simulation that takes into account connectivity and with the D-Wave QPU for $^2P_{1/2}$ and $^2P_{3/2}$ states for all the considered ions.}
    \label{fig:sim_hw}
\end{figure}

\begin{figure*} 
\begin{tabular}{ c c c c }
\includegraphics[scale=0.625]{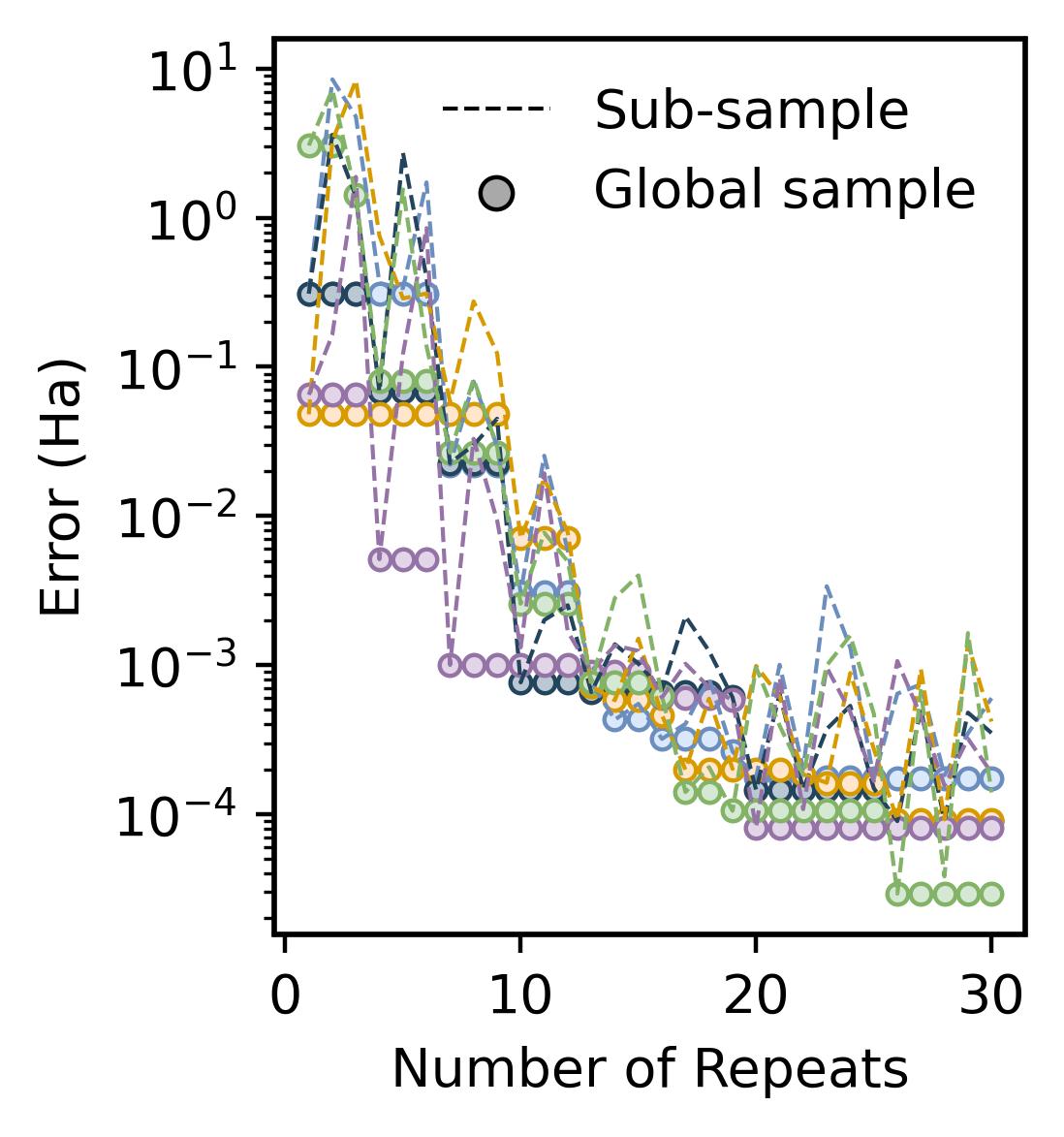} & \includegraphics[scale=0.625]{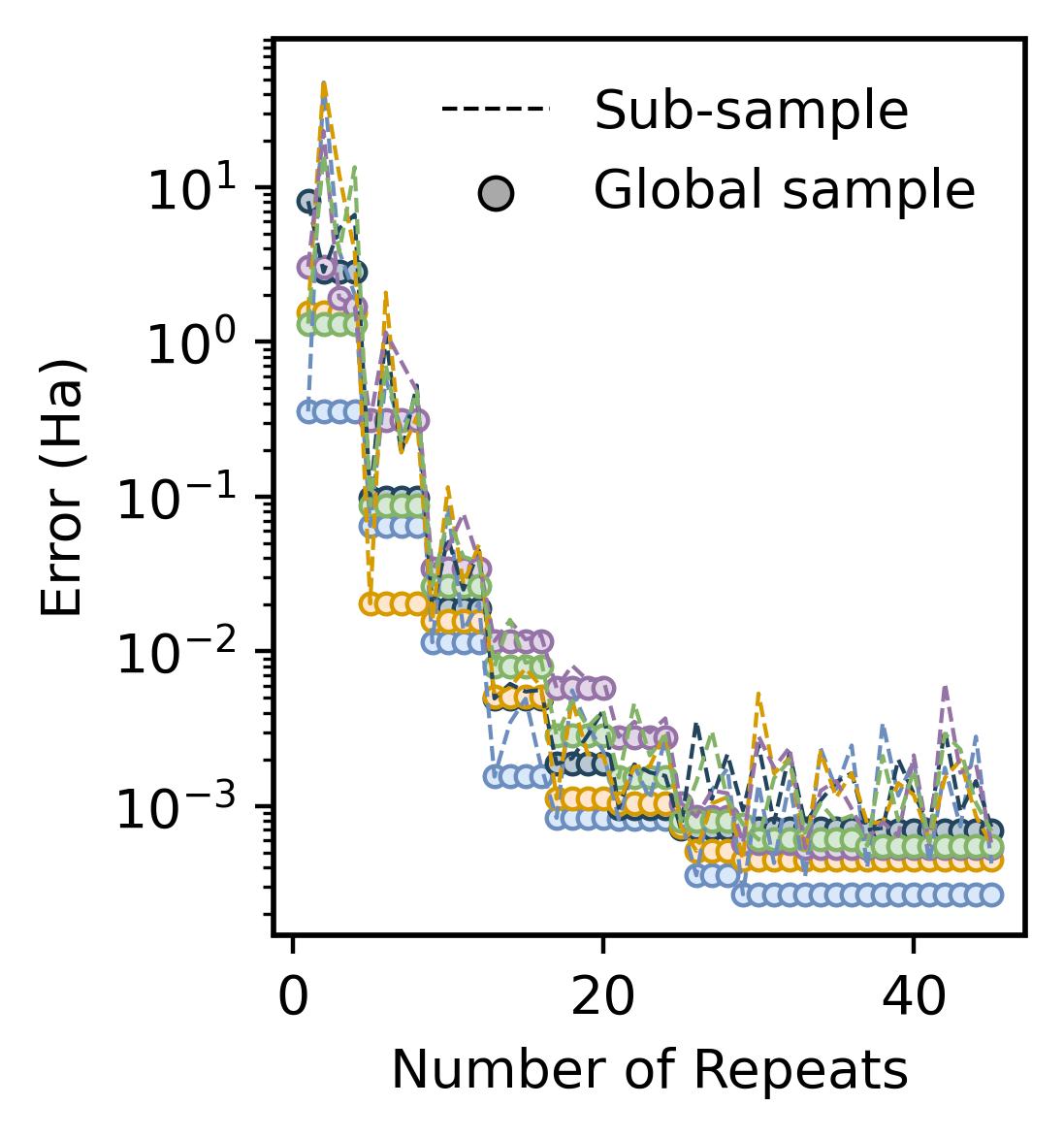} & \includegraphics[scale=0.625]{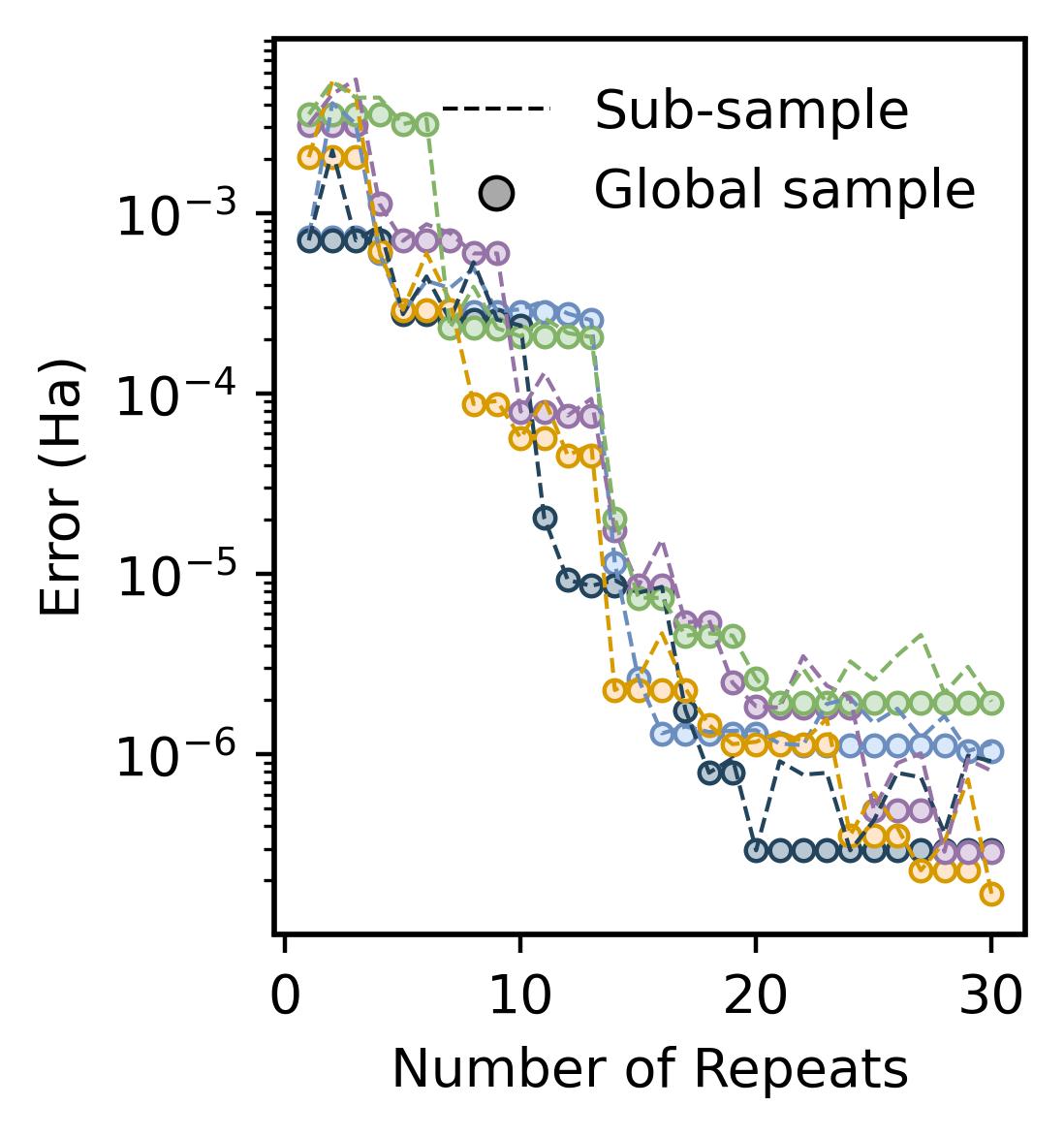} &\includegraphics[scale=0.625]{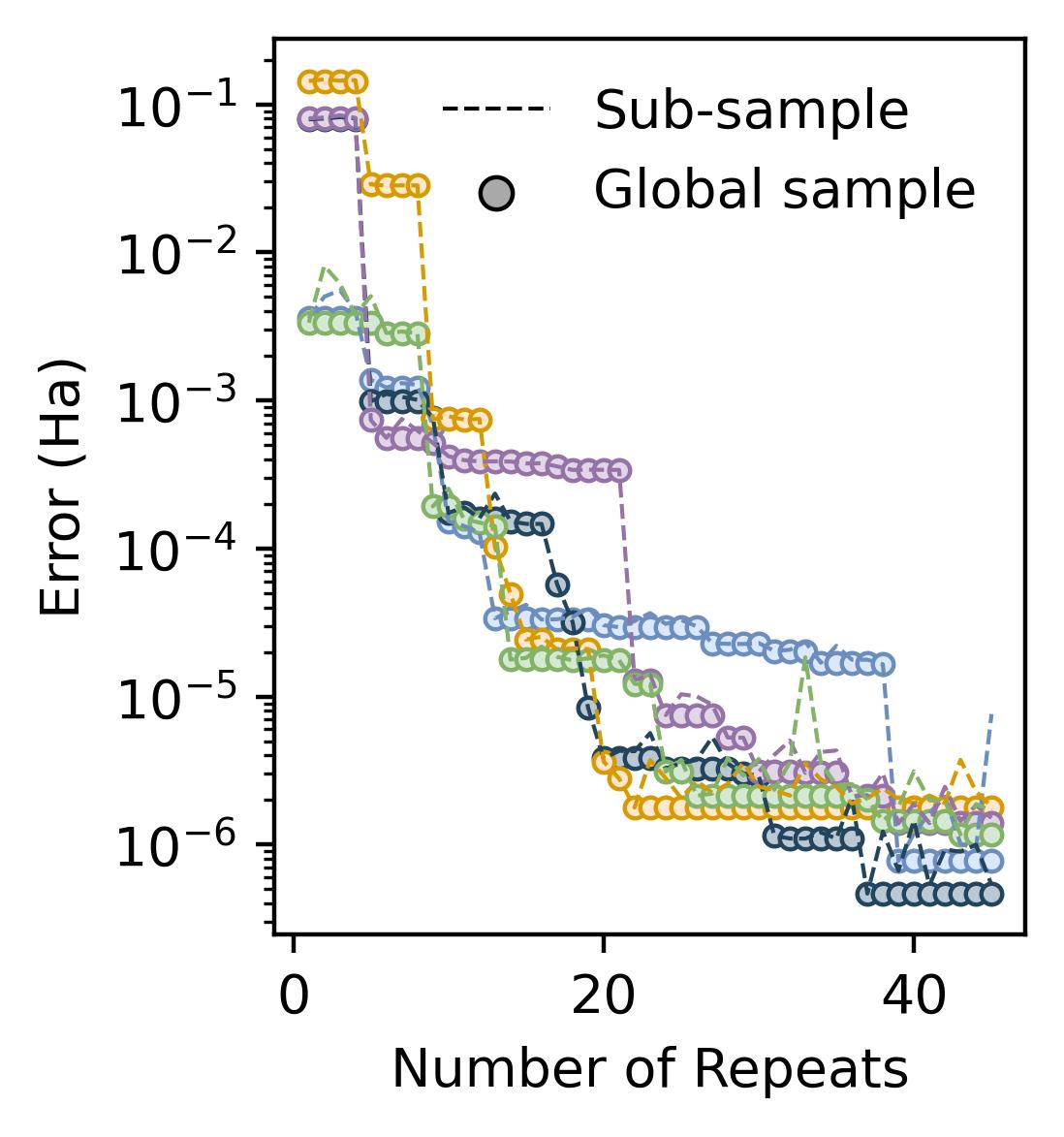} \\
\quad \quad \quad \, (a) & \quad \quad \quad \, (b) & \quad \quad \quad \, (c) & \quad \quad \quad \, (d) 

\end{tabular}
\caption{\label{fig:hw_vs_hwsd} Error in energy with respect to relCI versus number of Repeats in our workflow for $^2P_{1/2}$ (subfigure (a)) and $^2P_{3/2}$ (subfigure (b)) states of B-like Kr using the D-Wave Advantage QPU. Subfigures (c) and (d) provide the same data, but by employing the D-Wave Advantage QPU with steepest descent. The dashed lines present the error at every repeat, while circles present the error of the global sample. Different colour denotes different repetition of the experiment.
}
\end{figure*}

\begin{table} 
\label{MainTable}
\caption{\label{table-results} FSS values for boron-like ions. `relCI' refers to numerical relativistic CI calculations, `Simulation' and `Hardware' gives our mean (over five repetitions) of relativistic QAE performed on a traditional computer (with qubit connectivity taken into account) and D-Wave Advantage machine respectively. $\Delta_X\% = \frac{E_{\mathrm{relCI}}-E_{X}}{\mathrm{relCI}} \times 100$. `Expt' stands for the experimental value (in Ha). }
\begin{tabular}{ c c c c c }
    \hline \hline 
        System & relCI & Simulation  & Hardware  & Expt. \\
         &  &  ($\Delta_S\times 10^{-2}\%$) &  ($\Delta_H\times 10^{-2}\%$) &  \\
    \hline
        Ca$^{15+}$ &   0.163444  & 0.163513(-4.26) & 0.163673(-14.06) & 0.166831\\
        Fe$^{21+}$ &   0.532338  & 0.532527(-3.55) & 0.532572(-4.41) & 0.538863\\
        Kr$^{31+}$ &   2.228474  & 2.228836(-1.62) & 2.228881(-1.83) & 2.244283\\
        Mo$^{37+}$ &   4.368104  & 4.368713(-1.39) & 4.368535(-0.99) & 4.393976\\        
    \hline \hline 
\end{tabular}
\end{table}

\section{Results}
In Table~\ref{table-results}, we present our FSS results using QAE and compare with those of our numerical relativistic configuration interaction calculations~(RelCI) and experiments for Ca$^{15+}$, Fe$^{21+}$, Kr$^{31+}$ and Mo$^{37+}$. Our QAE computations were carried out both on traditional (simulation~\cite{simulatedAnnealing}) and quantum annealing (D-Wave Advantage) hardware. Our simulation results account for qubit connectivity. Our hardware results agree with the simulation and relCI computations to within  0.1 mHa. Finally, it is important to note that our FSS values agree with the experimental results to around 99 percent on the average.  

Noting that FSS involves finding energy differences and Table~\ref{table-results} presents only the former, we give the errors in the individual energies with respect to relCI energies, for all the boron-like ions that were considered, in Fig.~\ref{fig:sim_hw} using both simulation and hardware. The data corresponding to the individual energies can be found in Table~\ref{RC33-table} of the Supplementary Material. Our results from that figure show that our QAE approach can predict not only the FSS values but also the individual energies to within $2.6\times10^{-5}$ percentage. 
\begin{table}
\caption{\label{table_hw_hwsd}  Percentage fractional difference of energies of $^2P_{1/2}$ and $^2P_{3/2}$ states for B-like Kr from simulation~(S), hardware~(H), and hybrid sampler~(HS) with respect to relCI. Hybrid sampler here refers to computation performed on D-Wave Advantage with steepest descent optimizer.}
\begin{tabular}{ c c c c c c }
    \hline \hline 
         State & relCI (Ha) & $\Delta_S\times 10^{-5}\%$ & $\Delta_H\times 10^{-5}\%$ & $\Delta_{HS}\times 10^{-5}\%$ \\
    \hline
         $^2P_{1/2}$ & -1730.364613  & 0.70 & 0.53 & 4.30 $\times 10^{-3}$ \\
        $^2P_{3/2}$  & -1728.136140  & 2.80 & 2.90 & 5.42$\times 10^{-3}$ \\
        
    \hline \hline 
\end{tabular}
\end{table}

The accuracy of our results is due to the inclusion of the important physics and computational attributes that are appropriate for the evaluation of the FSS of atomic systems. This includes the addition of Breit interaction in the Hamiltonian, choosing the same set of optimised orbitals for the J=1/2 and J=3/2 states of boron-like ions, as well as workflow improvements such as our floating qubit encoding scheme, updating lambda iteratively with the information of energy from previous iterations, priority ordering of the expansion coefficients on the basis of perturbation theory, and finally small adjustments such as tuning the coefficient of the RMS of biases. The MCDF orbital optimization can be improved even further, thereby improving the atomic states, but nevertheless, it is useful for demonstrating the QAE approach with a larger configuration space. 

As a possible future extension, we could consider appending an optimization routine (for example, a discrete analogue of steepest descent) as a local search algorithm at the end of each repeat. We have considered a representative system, B-like Kr, and have carried out the aforementioned procedure, termed as `hybrid sampling’. Our results for the percentage fraction difference with respect to relCI are presented in Table~\ref{table_hw_hwsd}. Figure~\ref{fig:hw_vs_hwsd} presents a comparison of error (difference in the energy from relCI and energy from QAE algorithm) with number of Repeats for a representative system (boron-like Krypton), with subfigures (a) and (b) employing our QAE workflow and with (c) and (d) using the hybrid approach. Although the hybrid approach attains a given precision in lower number of Repeats, it involves additional computational cost. Our future directions could involve employing more sophisticated many-body methods in combination with hybrid workflows with other optimizers. 

We begin by seeking a precision of at least $10^{-3}$, that is, $E_{relCI}-E_{QAE}$ to be at least $10^{-3}$ with a high probability. We select $r$, the number of repeats, by scanning datasets for a given state, such as $^2P_{1/2}$, and repeating runs five times per atomic system. We pick the maximum $r$ where an error of $10^{-3}$ is reached in each of the twenty resulting curves and set $2r$ as the stopping criterion for hardware. The additional multiplicative factor, 2 is introduced keeping in mind the limitations of the current state-of-the-art hardware. For $^2P_{3/2}$ states, we use $1.5 r$ as the stopping criterion due to the larger matrix size and higher computation cost. Lastly, we chose the subQUBO sizes as an integral factor of the total QUBO size for convenience. Based on the above considerations, we chose $r = 30$ for $^2P_{1/2}$ states and $r = 45$ for $^2P_{3/2}$ states, with subQUBO sizes given in Table ~\ref{table-params_hardware}. In our work, we knew the relCI values. In the future when these values are unknown, we expect that this analysis can serve as an indicator for the stopping criteria in future computations with larger systems. The data is presented in Figures 2 and 3 in the supplementary material. 

\section{Conclusion}
In summary, we have developed the relativistic version of the QAE algorithm, which computes the minimum eigenvalues corresponding to specific atomic symmetries, and conducted a pilot study to calculate fine structure splittings in the boron-like atomic ions (Ca$^{15+}$, Fe$^{21+}$, Kr$^{31+}$, and Mo$^{37+}$). The method of eigenvalue estimation through quantum annealing may be of practical relevance for a diverse range of problems. A salient feature of our work involves improving the QAE workflow and includes an improved optimization strategy for the Lagrange multiplier occurring in the energy functional, priority listing of the CI coefficients, and floating qubit encoding method. With these improvements in place, we carried out relativistic calculations on both traditional computers as well as on quantum hardware. In the former, the simulations were carried out by accounting for the absence of all-to-all connectivity on the D-Wave devices. We find that our hardware results for fine structure splitting yield an average deviation of $1.1$ percent with respect to experiments and $2.71\times10^{-2}$ percent with respect to the relativistic CI calculations that we use to benchmark our results. We stress that our implementation allows us to accurately predict the individual energies of the states, thus enabling us to predict the fine structure splitting, which is an energy difference, accurately. To that end, we add that the individual energies themselves were evaluated to be within $\sim5\times10^{-5}$ percent of relativistic CI energies. Our work marks the first demonstration of relativistic many-body calculations carried out on quantum annealers, and our accurate results can be considered a stepping stone to future relativistic atomic and molecular calculations for novel applications, including probing new physics beyond the standard model of elementary particles. 

\begin{acknowledgements}
The classical computations were done on Rudra (SankhyaSutra Labs) supercomputers. The AWS Braket platform was used for cloud access to D-Wave Advantage 6.1 Hardware through the credits provided by the MeitY QCAL project (N-21/17/2020-NeGD, 2022-2024). K.S. acknowledges support from JST PRESTO ``Quantum Software" project (Grant No. JPMJPR1914).
\end{acknowledgements}


\bibliography{references, references1}

\onecolumngrid
\newpage
\section*{Supplementary Material}

\setcounter{equation}{0}
\setcounter{section}{0}
\setcounter{figure}{0}
\setcounter{table}{0}
\renewcommand{\thetable}{S\arabic{table}}
\renewcommand{\thefigure}{S\arabic{figure}}
\renewcommand{\theequation}{S\arabic{equation}}

\section{\label{supp_sec1}QAE Workflow}

We provide the details of the hybrid workflow which we use to minimize $F_Q$ in Eq.~\eqref{FQ} in the main text. The minimization is performed through a series of sub-space optimizations(defined as a `Repeat'), where each sub-space corresponds to a subset of coefficients to be optimized, and whose energy landscape is determined by a particular $\lambda$.

Fig.~\ref{fig:FIG1} in the main manuscript illustrates our QAE workflow. The first step consists of initializing all the coefficients, $\{a_{\alpha}\}$, to zero, and $\lambda$ as the CSF energy associated with the most dominant configuration, $\langle \Phi_0 | H| \Phi_0 \rangle$. This is followed by generating $F_Q$ and a priority list. The latter refers to a reordered list of the coefficients $\{a_\alpha\}$, generated from the input Hamiltonian matrix elements by considering that they are a first-order perturbation of the Dirac-Fock state.

The first step of a repeat is decomposition, which involves accounting for the lack of all-to-all connectivity and limited number of qubits available on real hardware. We select the first $\Gamma$ number of dominant coefficients from the sorted list, $\{a_\alpha\}_p \in \{a_\alpha\}$. $\Gamma$ is carefully chosen, so that the problem decomposes into as few subproblems as possible, while also ensuring that the results are not affected too much. The corresponding sub-problem of the QUBO $F_Q$ is solved by the annealing step, which returns 1000 samples (shots). Each shot provides a reduced set $\{q\}_r$. From that set, we calculate the reduced set $\{a_\alpha\}_r$, and we merge it with the other coefficients that were not in $\{a_\alpha\}_p$, and using it we obtain the minimum energy. For each of the subsequent repeats, $\lambda$ as well as the choice of coefficients are varied as explained next. 

We now explain the approach we have used for our choice of $\lambda$. Fig.~\ref{fig:EvsLam} in Section~\ref{supp_sec3} of the Supplementary Material shows the variation of the optimized energy with $\lambda$. Note that each data point in the figure, that is, the optimized energy for a given value of $\lambda$, is obtained by minimizing Eq~\eqref{ep1} on the classical optimizer. The QAE algorithm should ideally have the value $\lambda=\lambda_{opt}$, for which the optimized energy is minimum (and is equal to the relCI energy). In previous implementations of QAE, the optimal normalization penalty $\lambda_{opt}$ over an appropriate range of $\lambda$ is found by either scanning~\cite{Teplukhin2019CalculationAnnealer} or through bisection~\cite{Teplukhin2020ElectronicAnnealer}. In our work, as discussed earlier, we had chosen the CSF energy corresponding to the dominant configuration as the initial value for $\lambda$. Subsequently, for each repeat, we set the value of $\lambda$ to the energy obtained from the previous repeat. From CI method, we know that $\lambda_{opt}$ should be equal to $E_{relCI}$. This choice of the initial guess of $\lambda$ is based on physical grounds and requires fewer Repeats than the previously proposed scanning~\cite{Teplukhin2019CalculationAnnealer} and bisection~\cite{Teplukhin2020ElectronicAnnealer} methods. 

The strategy of subspace optimization involves solving a sub-QUBO of qubits corresponding to the most dominant coefficients in the priority list. The first Repeat selected the first $\Gamma$ dominant coefficients, allowing us to perform annealing for the sub-space with highest variation in the Energy. The next $\Gamma$ lower-order configurations are added successively in this way through the following Repeats. When all the coefficients are included in this way after a few Repeats, we start over again, thereby leading to further tuning of the coefficients. Note that during this method, $\lambda$, $\sigma$ and $\mu_\alpha$ are also varying according to their own procedure.

The second step is embedding, which embeds a densely connected QUBO to the Dwave QPU's topology by mapping some logical qubits to many hardware qubits and strongly coupling these hardware qubits together. This method, therefore, increases the number of qubits required. Here we focused on optimizing the strong coupling, called the chain strengths such that the chained qubits remain strongly coupled during quantum annealing, but also at the same time do not dominate the QUBO model’s energy. After testing different strategies in simulations, we found that using the RMS of the subproblem’s quadratic biases along with a multiplicative factor set between 0.75 and 1 gave the best results.

The annealing step consists of providing the sub-problem adapted to the D-Wave annealer's topology to the quantum annealer and obtaining a fixed number of samples each time (1000 for the present case). Note that the samples here mean the coefficients in the qubit space. The quantum annealer performs an adiabatic evolution from the transverse field Ising hamiltonian to the QUBO Hamiltonian given in Eq.~\eqref{FQ}~\cite{Johnson2011QA}. 

The composition and post-processing steps involve merging the obtained samples with the main sample, back-converting them to the expansion coefficients, scaling them individually and finally comparing with the known best solution from previous Repeats. The main sample refers to the solution of the complete QUBO problem (initialized to zeros). Scaling here refers to scaling $\{a_\alpha\}$ such that their absolute maximum becomes -1. This allows us to increase the smaller coefficients so that they can be captured with fewer qubits per coefficient~(K), and deters the iterative optimization to navigate toward trivial wavefunction solutions. The scaling step does not degrade the result because the normalization condition is not strictly imposed but rather with the addition of a penalty.

After each Repeat, the subset of coefficients, $\sigma$, $\mu_\alpha$, and $\lambda$ are changed in a systematic manner, and after several Repeats, we select the lowest energy among all Repeats.
\section{\label{secII_qubo_deri}The expression for QUBO form of the energy functional}

The energy functional to be minimized is given by
\begin{equation}
    \epsilon = \sum_{\alpha, \beta = 0}^{\mathcal{B} - 1} a_\alpha a_\beta(H_{\alpha\beta}-\lambda\delta_{\alpha\beta}).
\end{equation}
where $\{a\}$ are the expansion coefficients, H is the DCB Hamiltonian and $\lambda$ is the Lagrange multiplier.\\
We use the floating qubit encoding:
\begin{equation}
    a_\alpha = \sigma \sum_{k=0}^{K-1}f_k 2^{-k}q_k^\alpha + \mu_\alpha 
\end{equation}
where $f_k$ is -1 if $k=0$ otherwise 1, and $\{\mu\}$ and $\{\sigma\}$ are constants determined at each iteration.\\
The energy functional ($\epsilon$) then becomes 


\begin{eqnarray}
    \epsilon &=& \sum_{\alpha, \beta=0}^{\mathcal{B} - 1} \left( \sigma \sum_{n=0}^{K-1} f_n 2^{-n}q_n^\alpha + \mu_\alpha \right)\left( \sigma \sum_{m=0}^{K-1} f_m 2^{-m}q_m^\beta + \mu_\beta \right) \left(H_{\alpha\beta}-\lambda\delta_{\alpha\beta}\right)  \nonumber  \\
    &=& \sum_{\alpha, \beta = 0}^{\mathcal{B} - 1}\left(\sigma^2 \sum_{n, m =0}^{K-1}f_n f_m 2^{-(n+m)}q_n^\alpha q_m^\beta +\mu_\beta \sigma \sum_{n=0}^{K-1}f_n 2^{-n}q_n^\alpha + \mu_\alpha \sigma \sum_{m=0}^{K-1}f_m 2^{-m}q_m^\beta + \mu_\alpha\mu_\beta \right)H_{\alpha \beta}' \nonumber \\
    &=& \sum_{\alpha, \beta = 0}^{\mathcal{B} - 1}\left( \sigma^2 \sum_{n, m =0}^{K-1}f_n f_m 2^{-(n+m)}q_n^\alpha q_m^\beta +2\mu_\beta \sigma \sum_{n=0}^{K-1}f_n 2^{-n}q_n^\alpha + \mu_\alpha\mu_\beta \right)H_{\alpha \beta}' \nonumber \\
    &=& \sum_{\alpha, \beta = 0}^{\mathcal{B} - 1}\left( \sum_{n,m = 0}^{K-1}q_n^\alpha A q_m^\beta + \sum_{n = 0}^{K-1} B q_n^\alpha + C\right) 
\end{eqnarray}
where $\delta_{nm}$ is 1 if $n=m$ and 0 otherwise, and 
\begin{eqnarray}
A &=& (\sigma^2 f_n f_m 2^{-(n+m)})H_{\alpha \beta}', \nonumber \\ 
B &=& (2\mu_\beta \sigma f_n 2^{-n})H_{\alpha \beta}', \nonumber \\ 
C &=& \mu_\alpha \mu_\beta H_{\alpha \beta}', 
\text{ and } \nonumber \\ 
H_{\alpha \beta}' &=& (H_{\alpha \beta}-\lambda   \delta_{\alpha \beta}). \nonumber
\end{eqnarray}

\section{\label{supp_sec3}Additional Data}

\begin{figure}[!h]
    \centering
    \includegraphics[scale=0.45]{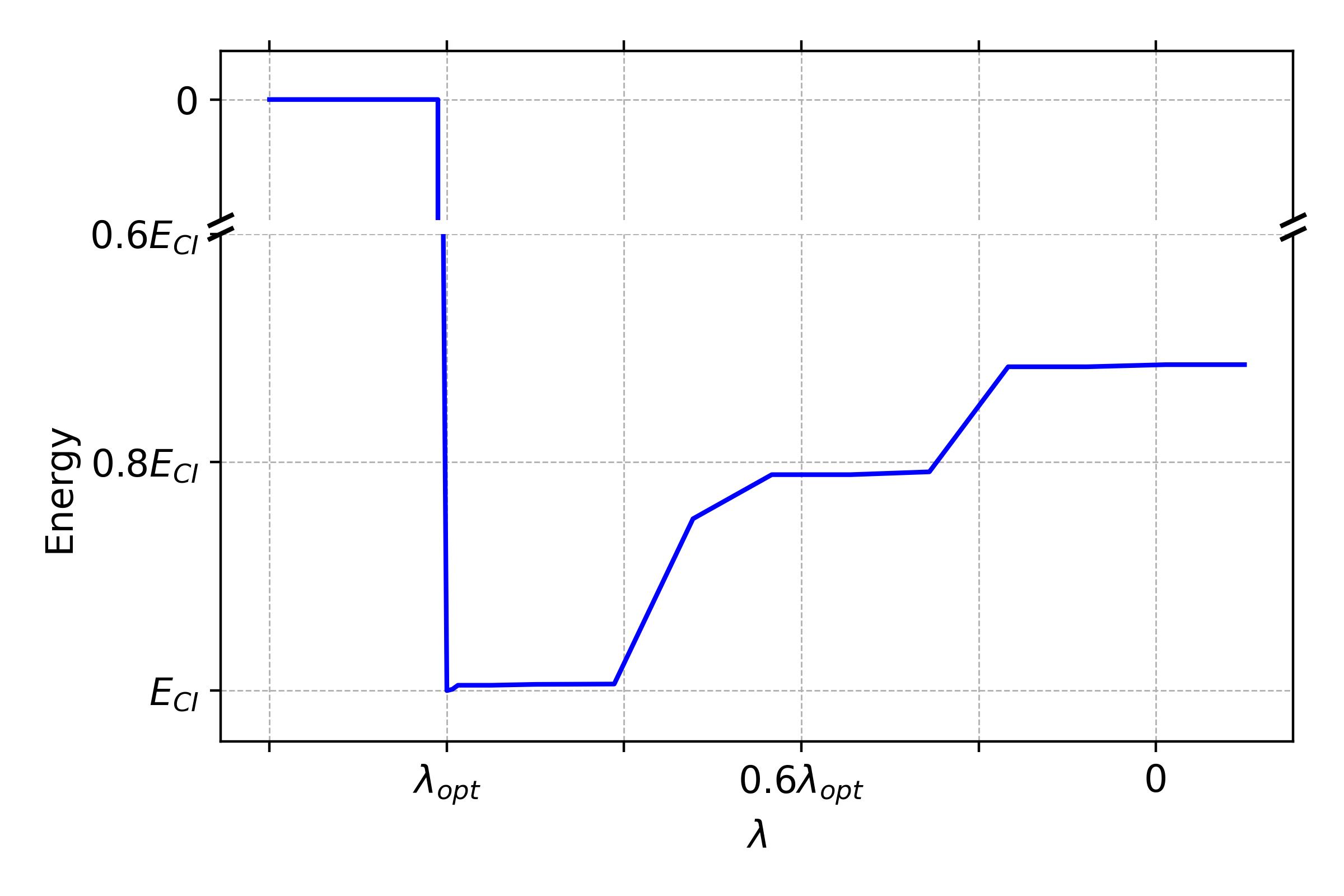}
    \caption{Variation of energy, $E$, with $\lambda$, while minimizing the energy functional, $\epsilon$. }
    \label{fig:EvsLam}
\end{figure}

\begin{table}[!h]
\caption{\label{S_ED}Comparison of results from QAE with default decomposition method against QAE with perturbation theory-inspired decomposition method, for boron-like Krypton. $\Delta$ and $\sigma$ refer to mean and standard deviation errors respectively, from exact diagonalization over ten runs. }

\begin{tabular}{l|c ccc|cccc}
\hline \hline
\multicolumn{1}{c|}{} & \multicolumn{4}{c|}{$^2P_{1/2}$}                     & \multicolumn{4}{c}{$^2P_{3/2}$}                          \\ \hline
\multicolumn{1}{c|}{} & Energy (Ha)      & $\Delta\times10^{-4}$\hspace{0.1cm}    & $\sigma\times10^{-4}$\hspace{0.1cm}  & Repeats & Energy        & $\Delta\times10^{-4}$\hspace{0.1cm} & $\sigma\times10^{-4}$\hspace{0.1cm} & Repeats \\ \hline
Perturbation-based \hspace{0.1cm}  & -1730.364181 & 0.71 & 0.37  & 30       & -1728.135698 & 4.42    & 1.89    &  45       \\ \hline
Default              & -1730.364543 & 4.32 & 2.72  & 53         & -1728.135312 & 8.27    & 3.06    & 73         \\ \hline \hline
\end{tabular}
\end{table}

\begin{table}
\caption{\label{table-params_hardware} Workflow parameters for our QAE computations. We note that SubQUBO refers to the size of $\Gamma$ after encoding.  }

\begin{tabular}{ c c c c }
    \hline \hline
        \quad   State \quad  & \quad  QUBO \quad  &  SubQUBO  &  Total  \\
         & size & size & Repeats\\
    \hline
        \quad  $^2P_{1/2}$ \quad  & 90 & 30 & 30 \\
         \quad  $^2P_{3/2}$ \quad  & 160 & 40 & 45 \\
    \hline \hline
\end{tabular}
\end{table}

\begin{table}[!h]
\caption{\label{RC33-table} Individual energies of chosen systems for $^2P_{1/2}$ and $^2P_{3/2}$ states obtained using relCI, Simulation(S) and Hardware(H), and their corresponding standard deviation. In the Table, `relCI' refers to the results obtained from relativistic configuration interaction (in Ha), `Simulation' gives our mean relativistic QAE results from a traditional computer (over five repetitions), but with hardware aspects such as connectivity taken into account, while the column `Hardware' gives the mean QAE results on the D-Wave Advantage machine (over five repetitions). SD refers to standard deviation.}
\begin{tabularx}{0.75\textwidth}{ >{\centering\arraybackslash}X >{\centering\arraybackslash}X >{\centering\arraybackslash}X >{\centering\arraybackslash}X >{\centering\arraybackslash}X >{\centering\arraybackslash}X }
    \hline \hline 
         & & Ca$^{15+}$ & Fe$^{21+}$ & Kr$^{31+}$ &  Mo$^{37+}$ \\
    \hline
        $^2P_{1/2}$ & relCI & -507.679395 & -878.442367 & -1730.364613 & -2387.586842  \\
                    & Simulation & -507.679333 & -878.442308 & -1730.364491 & -2387.586714  \\
                     & SD (S) & 4.93E-05 & 3.17E-05 & 3.18E-05 & 2.93E-05   \\
                    & Hardware & -507.679348 & -878.442291 & -1730.364520 & -2387.586660\\
                    & SD (H) & 1.61E-05 & 4.86E-05 & 4.61E-05 & 8.04E-05 \\ \hline   
        $^2P_{3/2}$ & relCI & -507.515951 & -877.910029 & -1728.136140 & -2383.218738   \\
                    & Simulation & -507.515819 & -877.909781 & -1728.135656 & -2383.218002  \\
                    & SD (S) & 5.76E-05 & 9.63E-05 & 1.67E-04 & 1.74E-04 \\
                    & Hardware & -507.515675 & -877.909719 & -1728.135639 & -2383.218125 \\
                    & SD (H) & 9.24E-05 & 1.01E-04 & 1.41E-04 & 1.86E-04\\ 
    \hline \hline 
\end{tabularx}
\end{table}

\begin{figure*} 
\begin{tabular}{ c c }
\quad \quad \quad \,\includegraphics[scale=0.7]{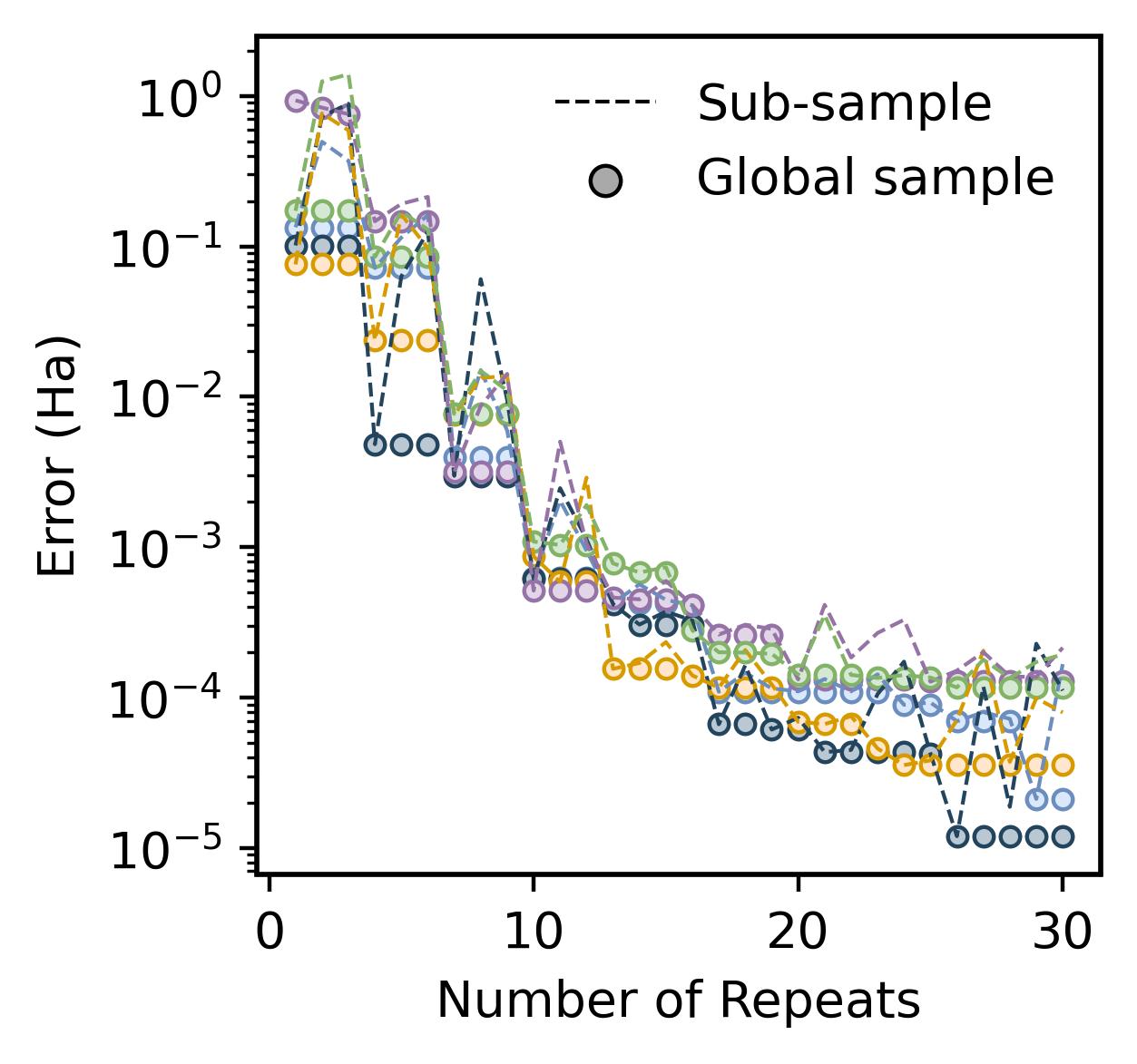} \quad \quad \quad \, & \quad \quad \quad \, \includegraphics[scale=0.7]{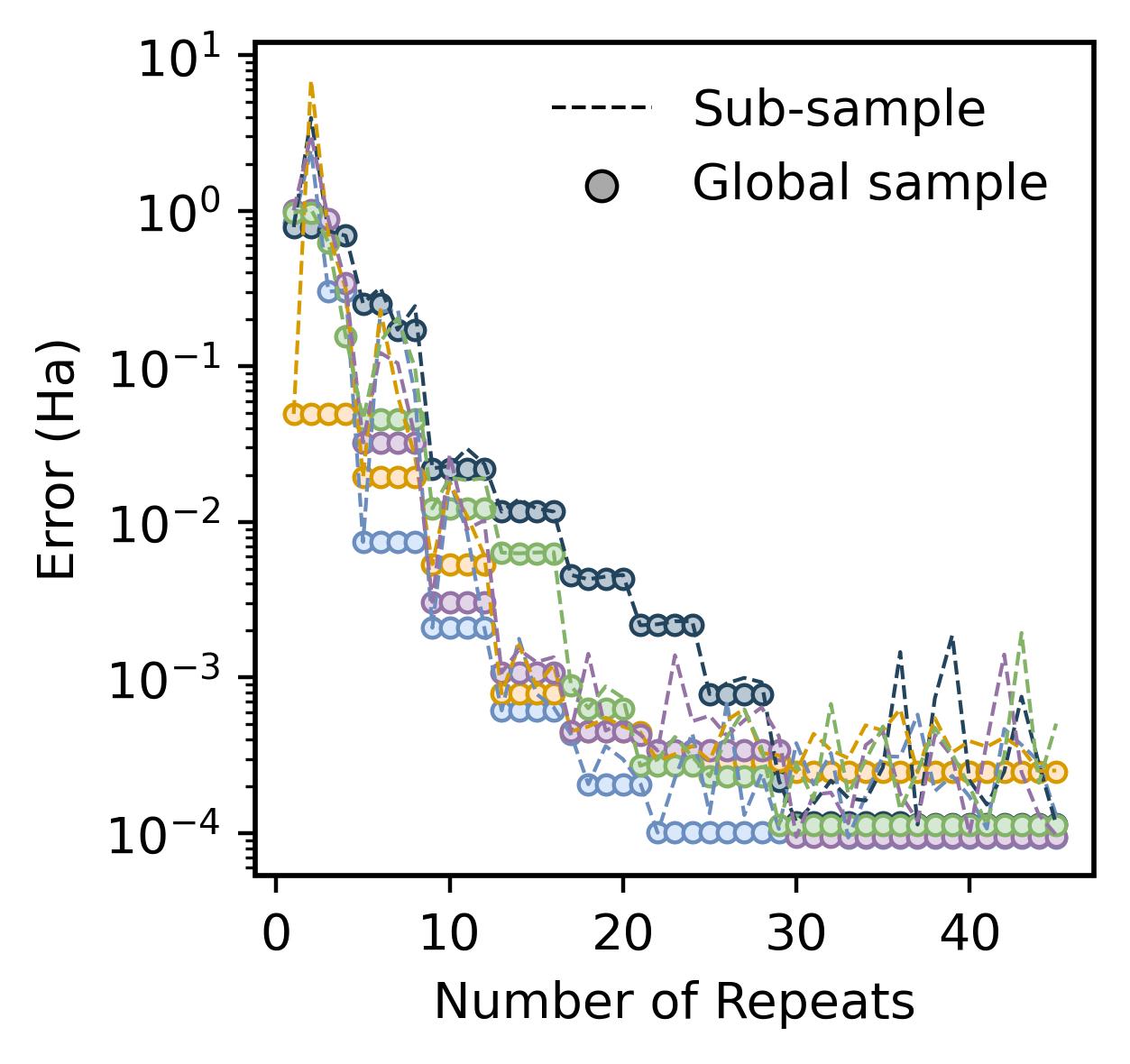} \quad \quad \quad \, \\
\quad \quad \quad \, (a) & \quad \quad \quad \, (b) \\
\includegraphics[scale=0.7]{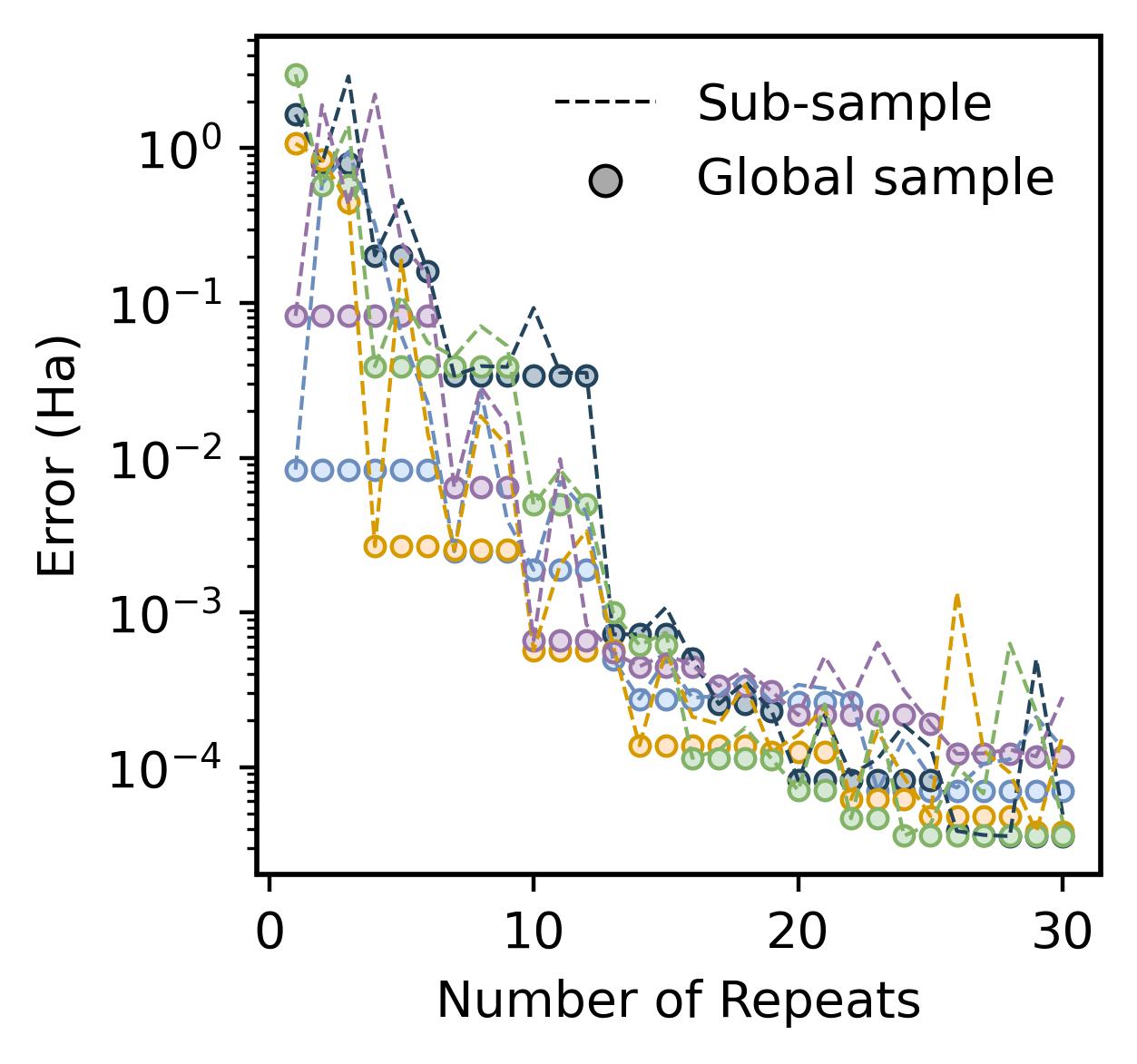} & \includegraphics[scale=0.7]{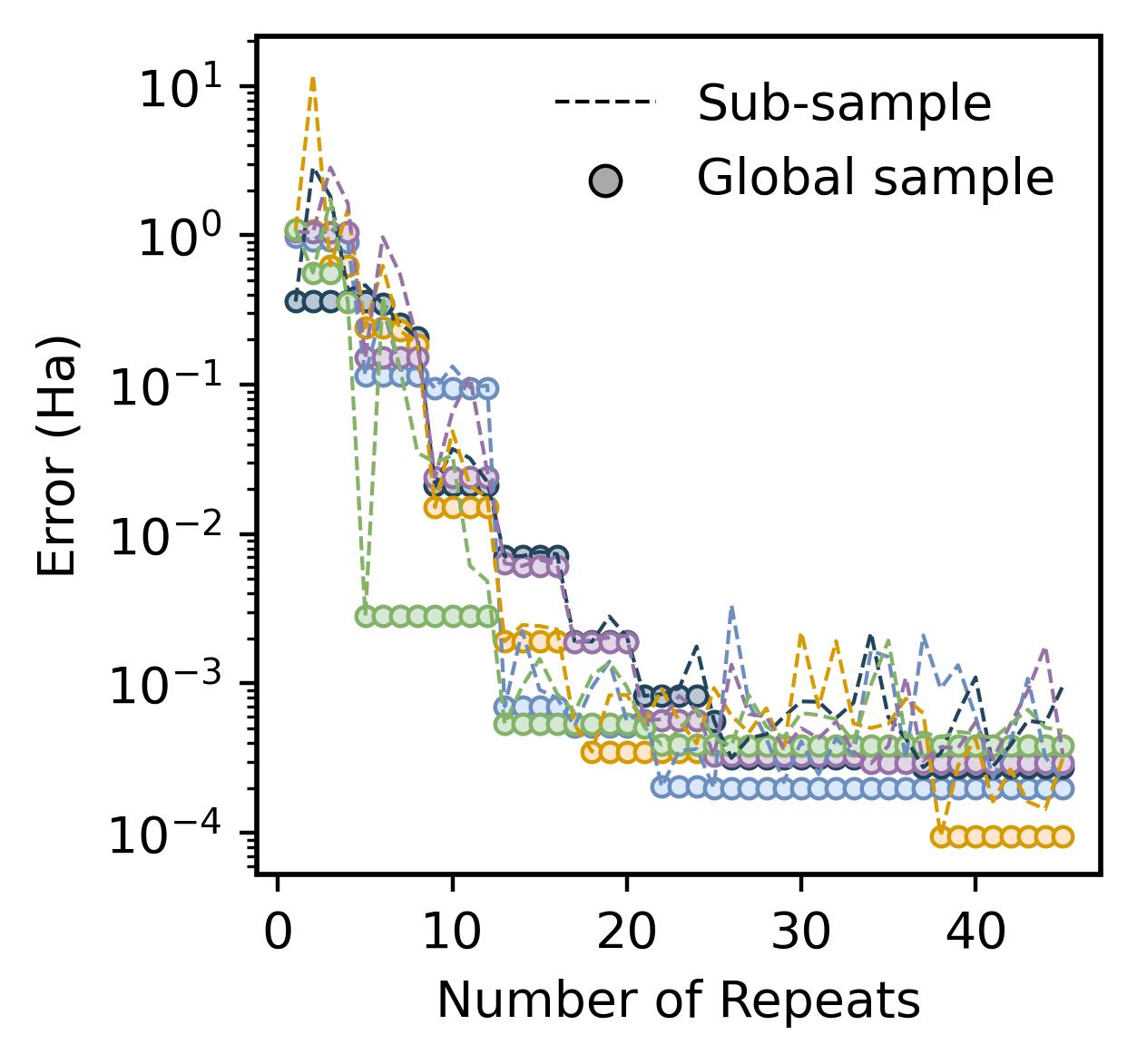}  \\
\quad \quad \quad \, (c) & \quad \quad \quad \, (d) \\
\includegraphics[scale=0.7]{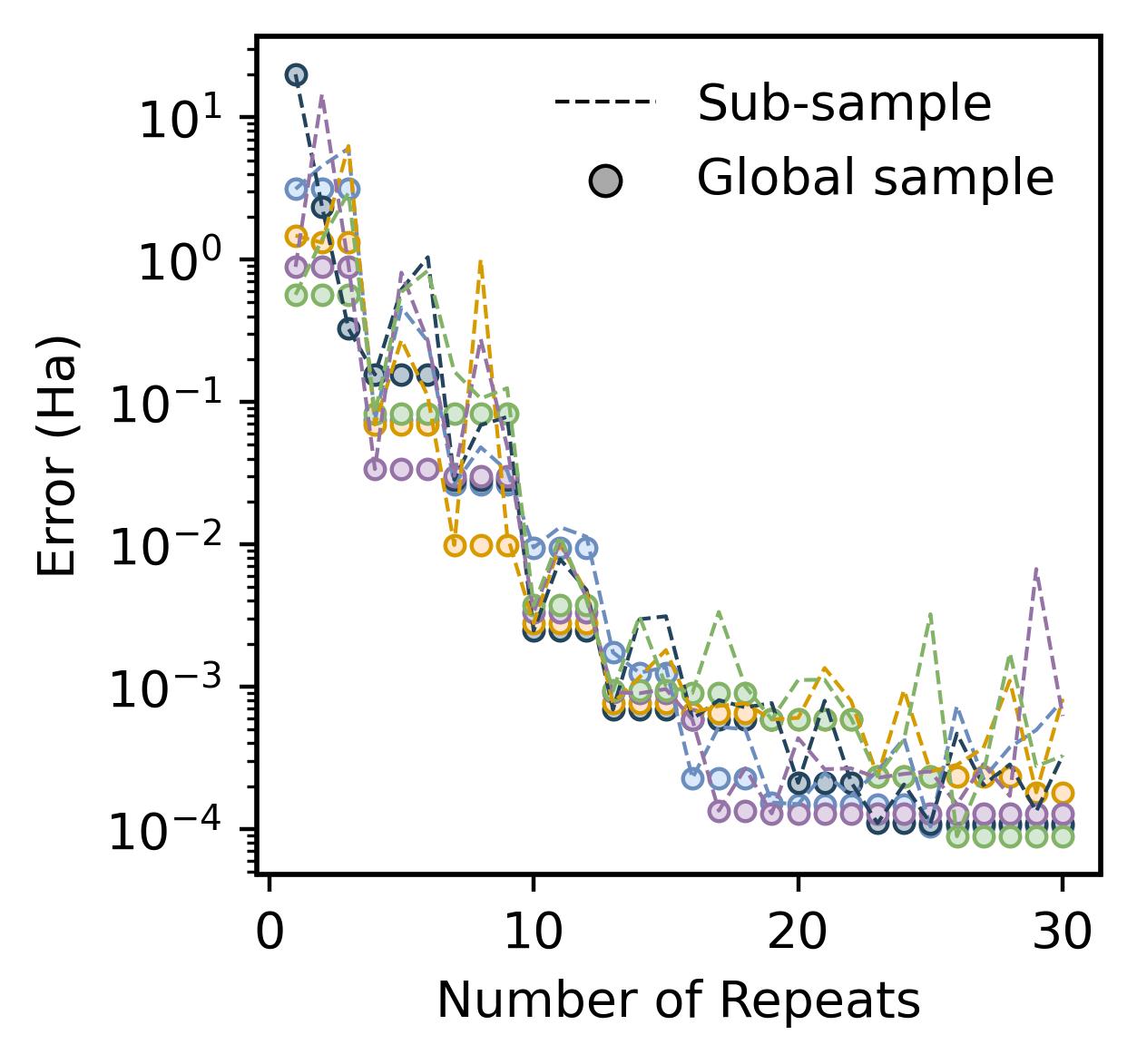} & \includegraphics[scale=0.7]{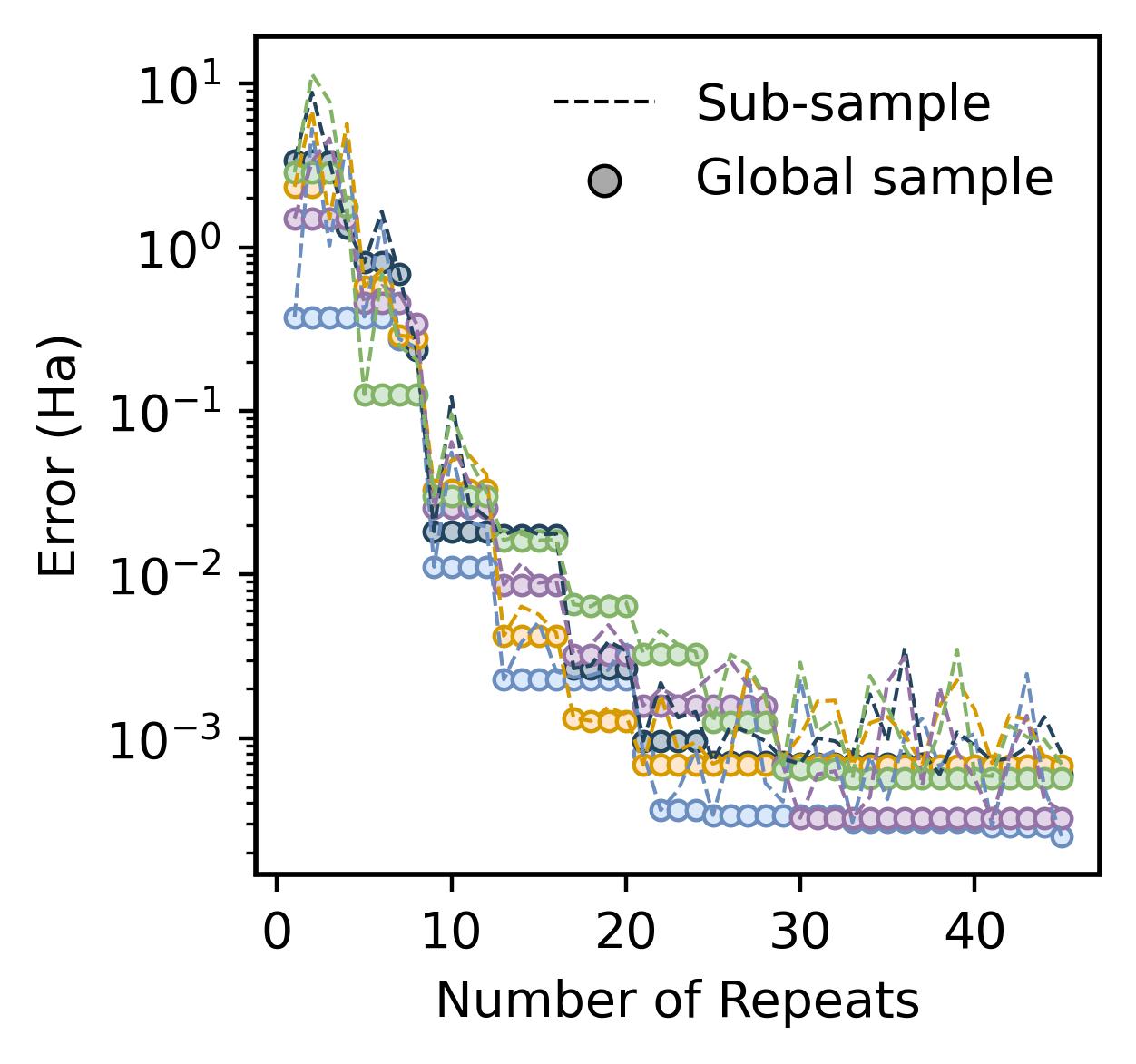}  \\
\quad \quad \quad \, (e) & \quad \quad \quad \, (f) \\
\includegraphics[scale=0.7]{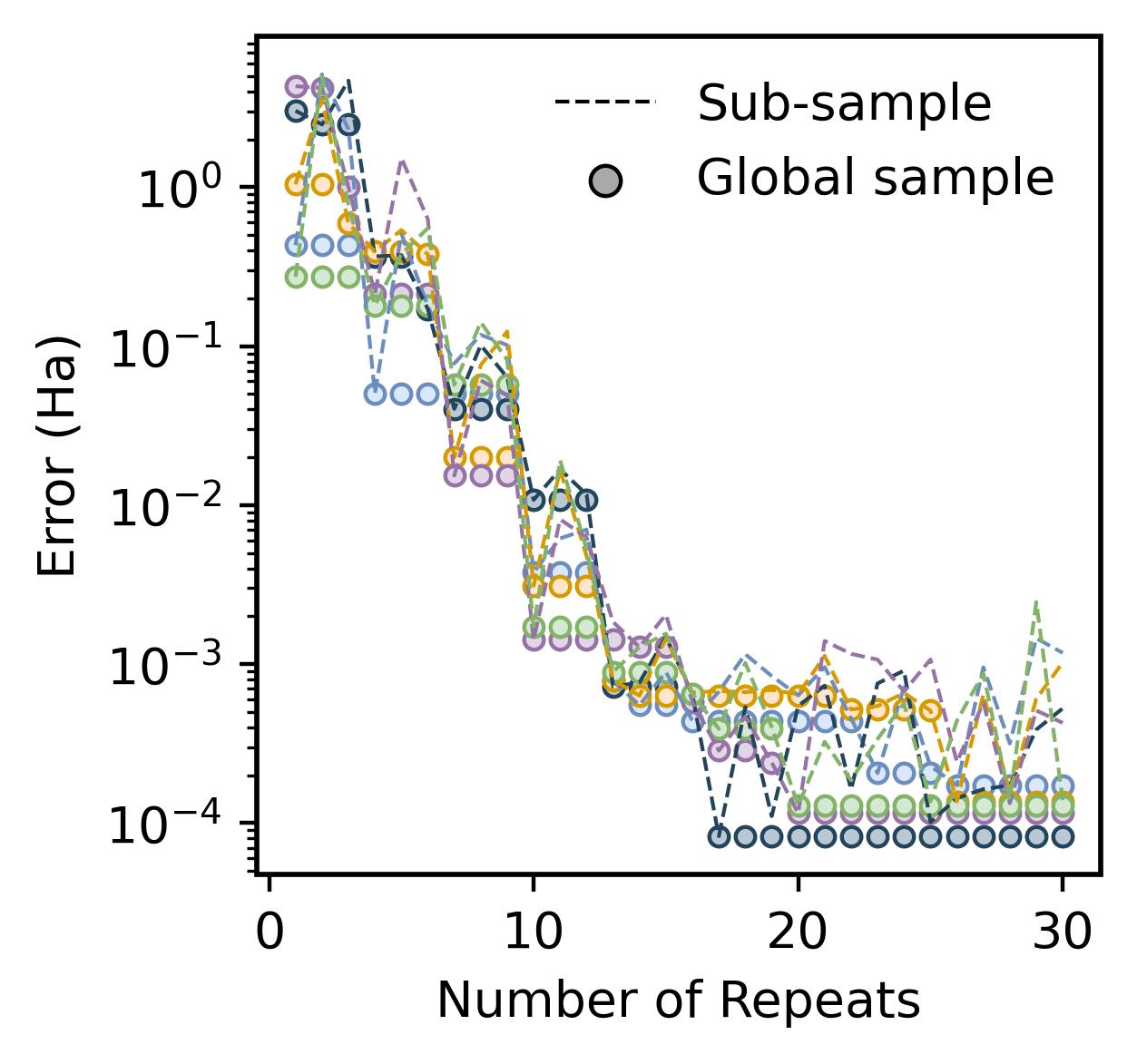} & \includegraphics[scale=0.7]{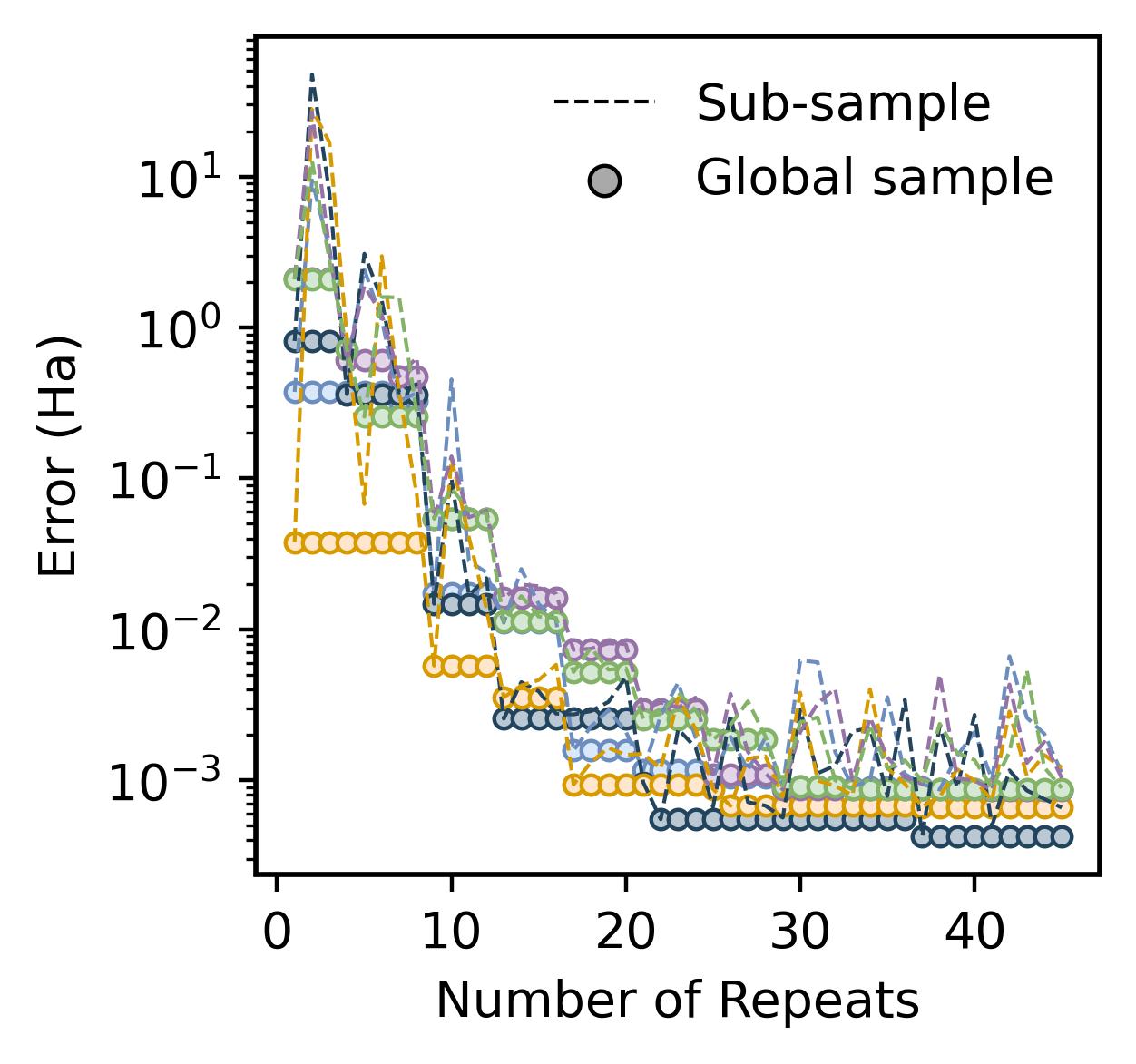}  \\
\quad \quad \quad \, (g) & \quad \quad \quad \, (h) \\

\end{tabular}
\caption{\label{fig:simallgraph} Error in energy with respect to relCI versus number of Repeats in our workflow for $^2P_{1/2}$ (subfigures (a), (c), (d), and (f)) and $^2P_{3/2}$ (subfigures (b), (d), (f), and (h)) states of all the considered boron-like atoms using the simulation, taking hardware architecture into consideration. The dashed lines present the error at every repeat, while circles denote the error in the global sample. Different colours denote different repetitions of the experiment. Subfigures ((a),(b)), ((c),(d)), ((e),(f)) and ((g),(h)) correspond to boron-like Ca, Fe, Kr and Mo respectively.
}
\end{figure*}

\begin{figure*} 
\begin{tabular}{ c c }
\quad \quad \quad \,\includegraphics[scale=0.7]{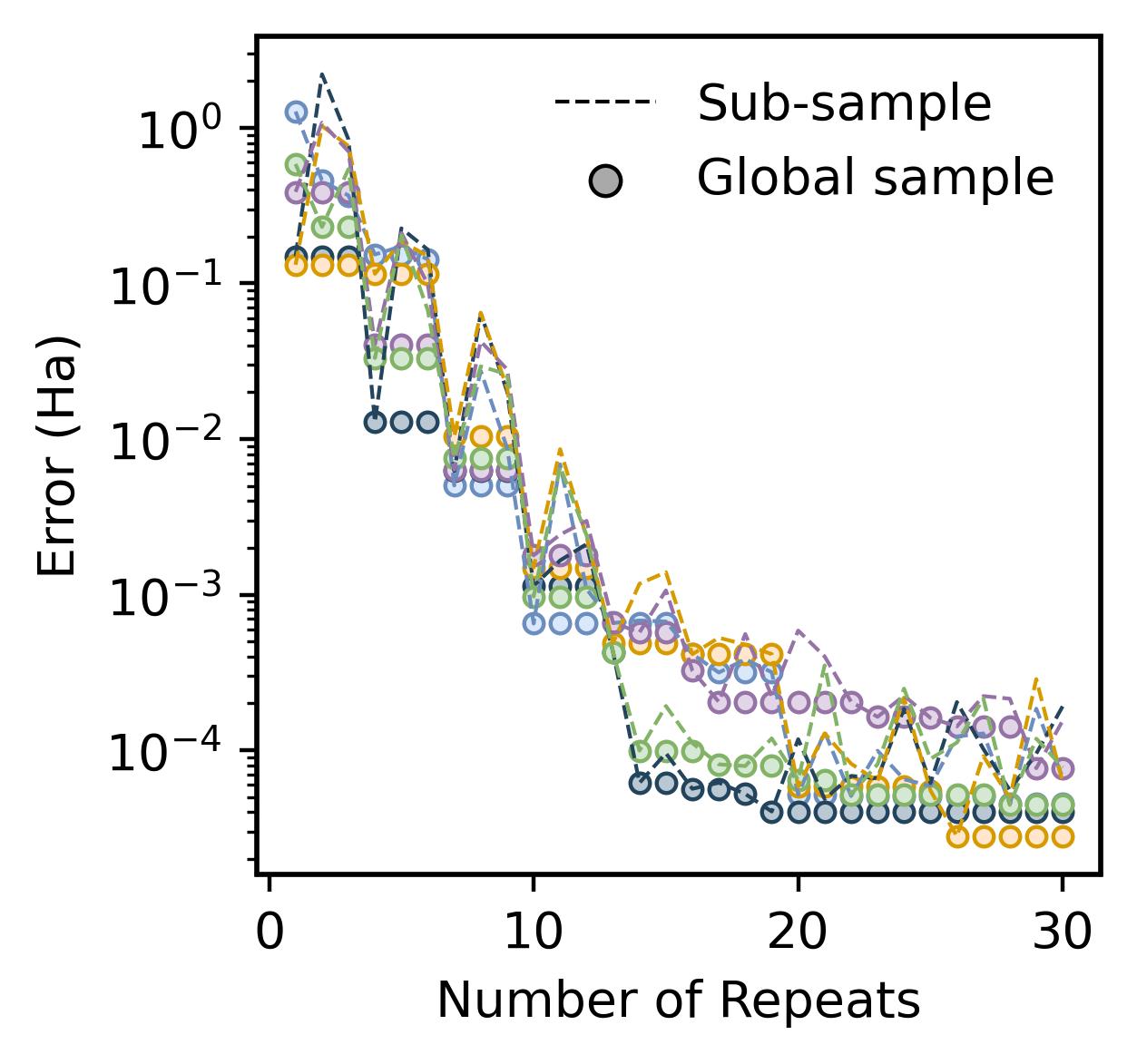} \quad \quad \quad \, & \quad \quad \quad \, \includegraphics[scale=0.7]{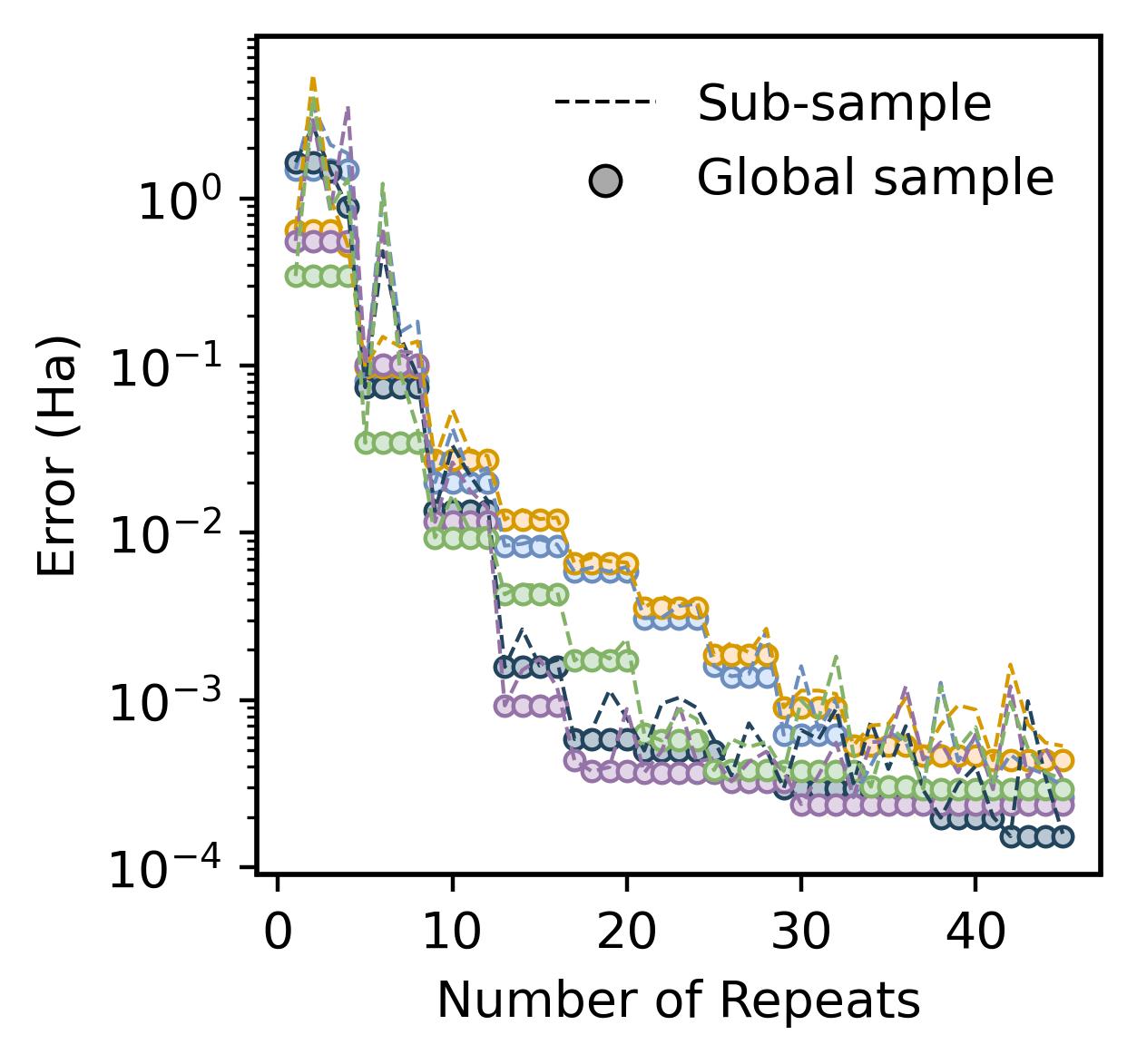} \quad \quad \quad \, \\
\quad \quad \quad \, (a) & \quad \quad \quad \, (b) \\
\includegraphics[scale=0.7]{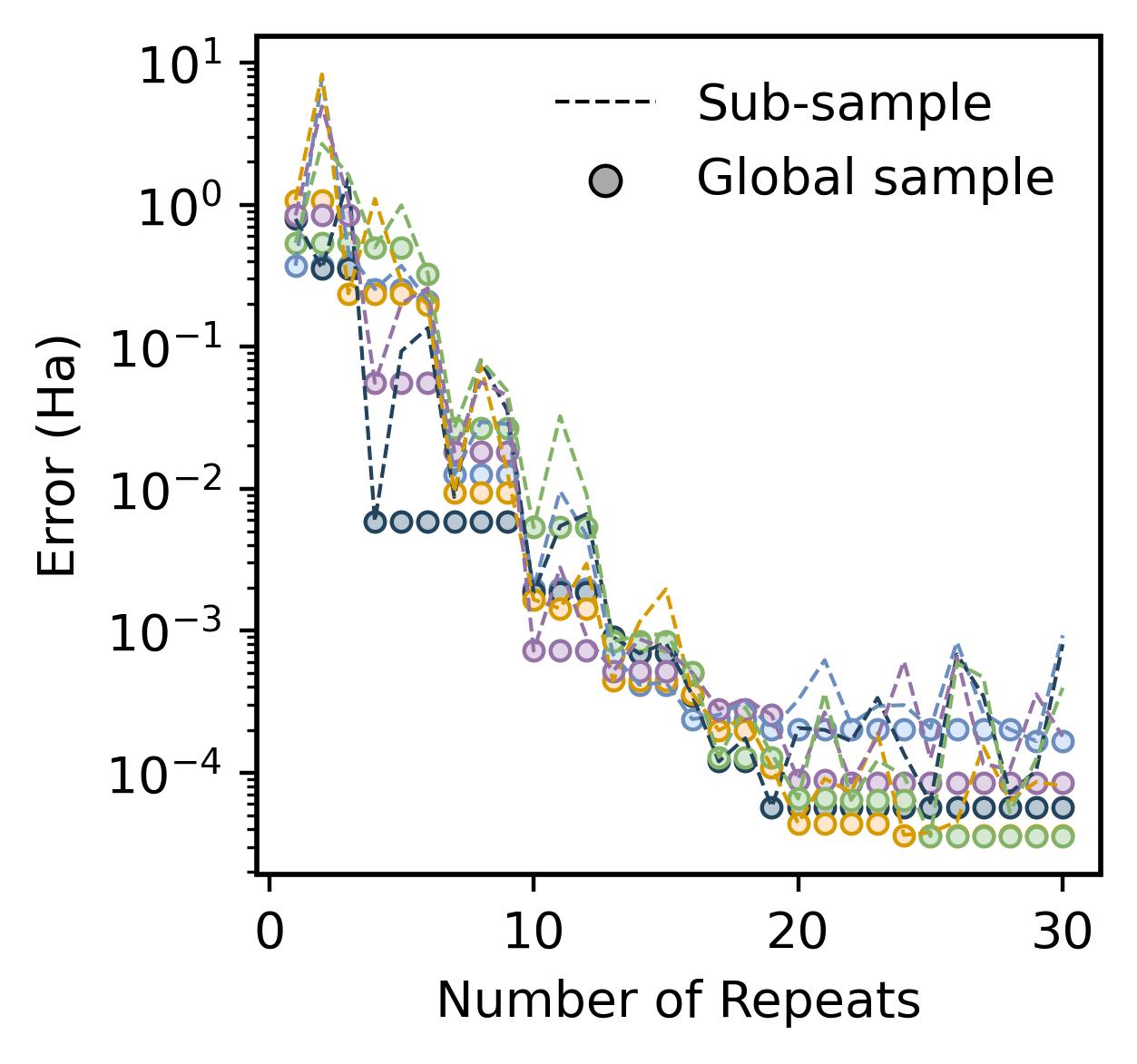} & \includegraphics[scale=0.7]{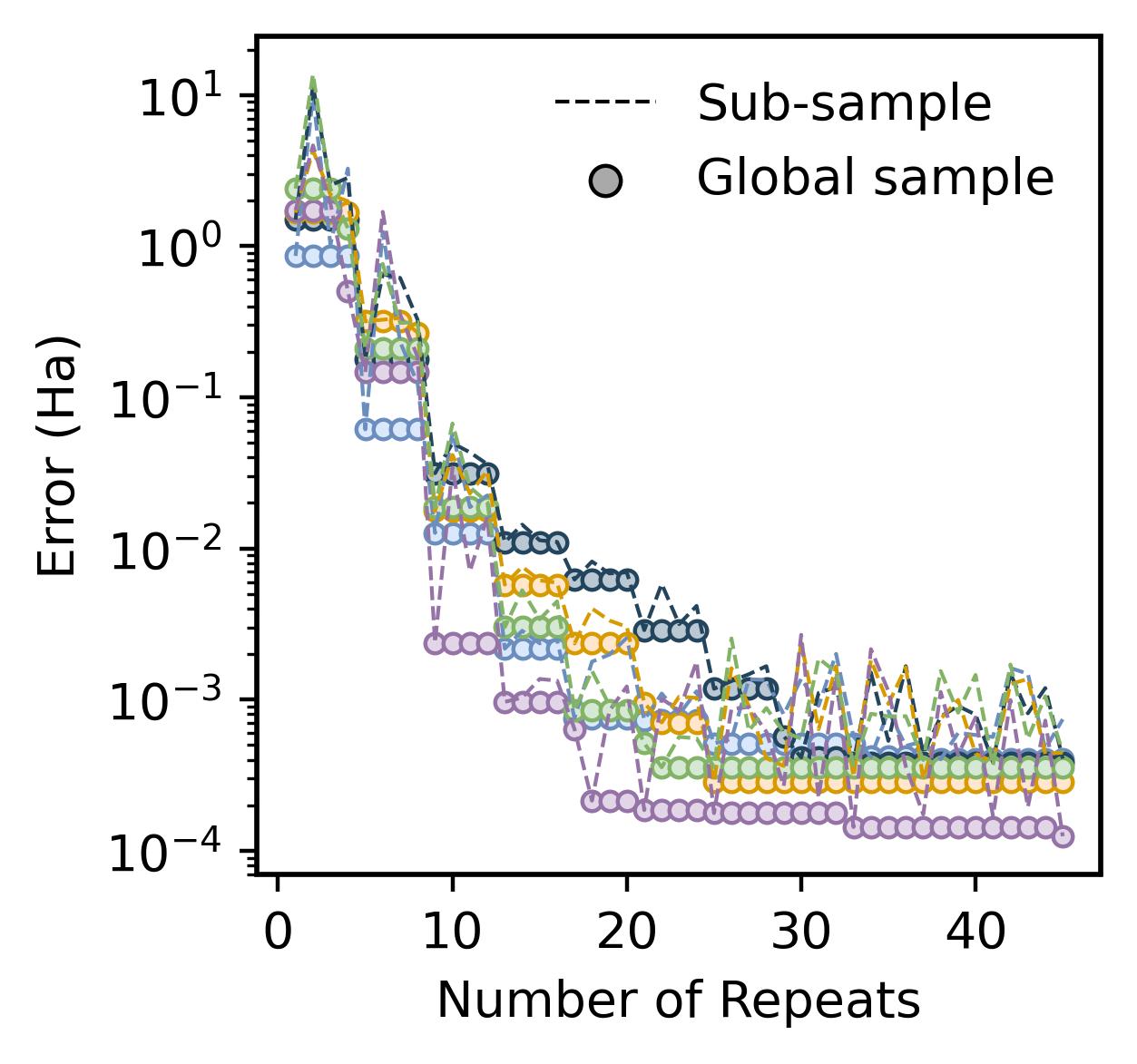}  \\
\quad \quad \quad \, (c) & \quad \quad \quad \, (d) \\
\includegraphics[scale=0.7]{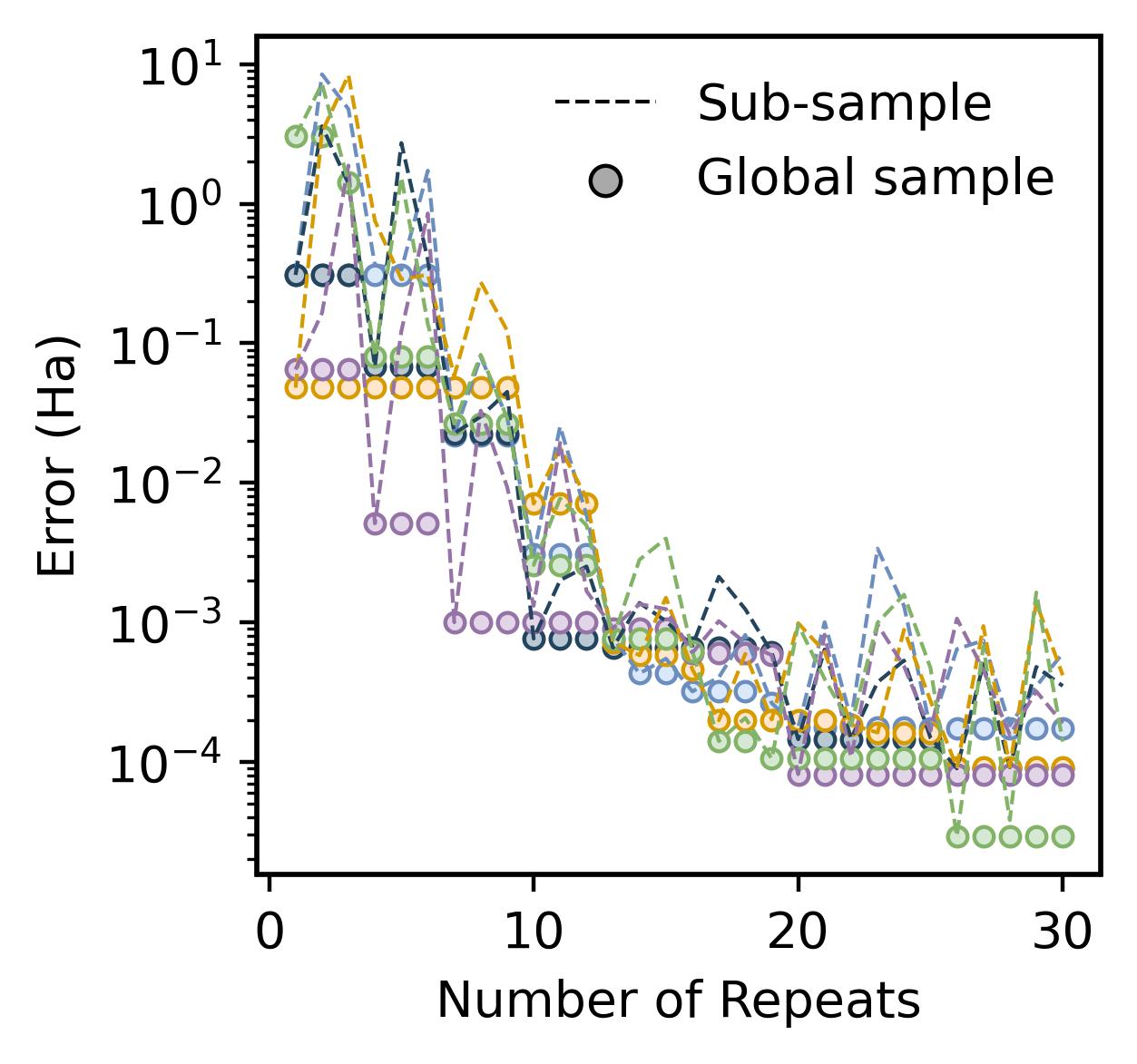} & \includegraphics[scale=0.7]{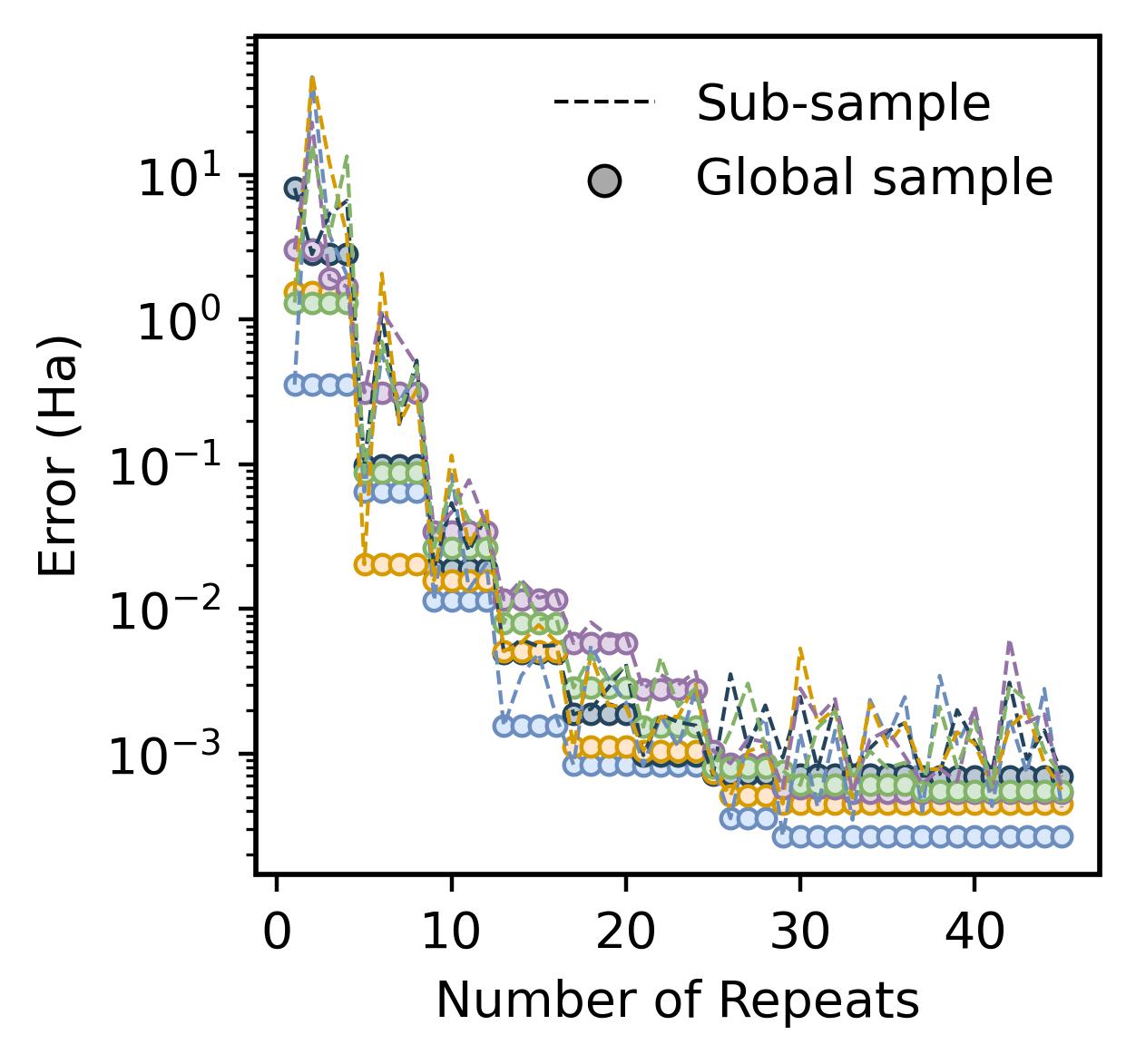}  \\
\quad \quad \quad \, (e) & \quad \quad \quad \, (f) \\
\includegraphics[scale=0.7]{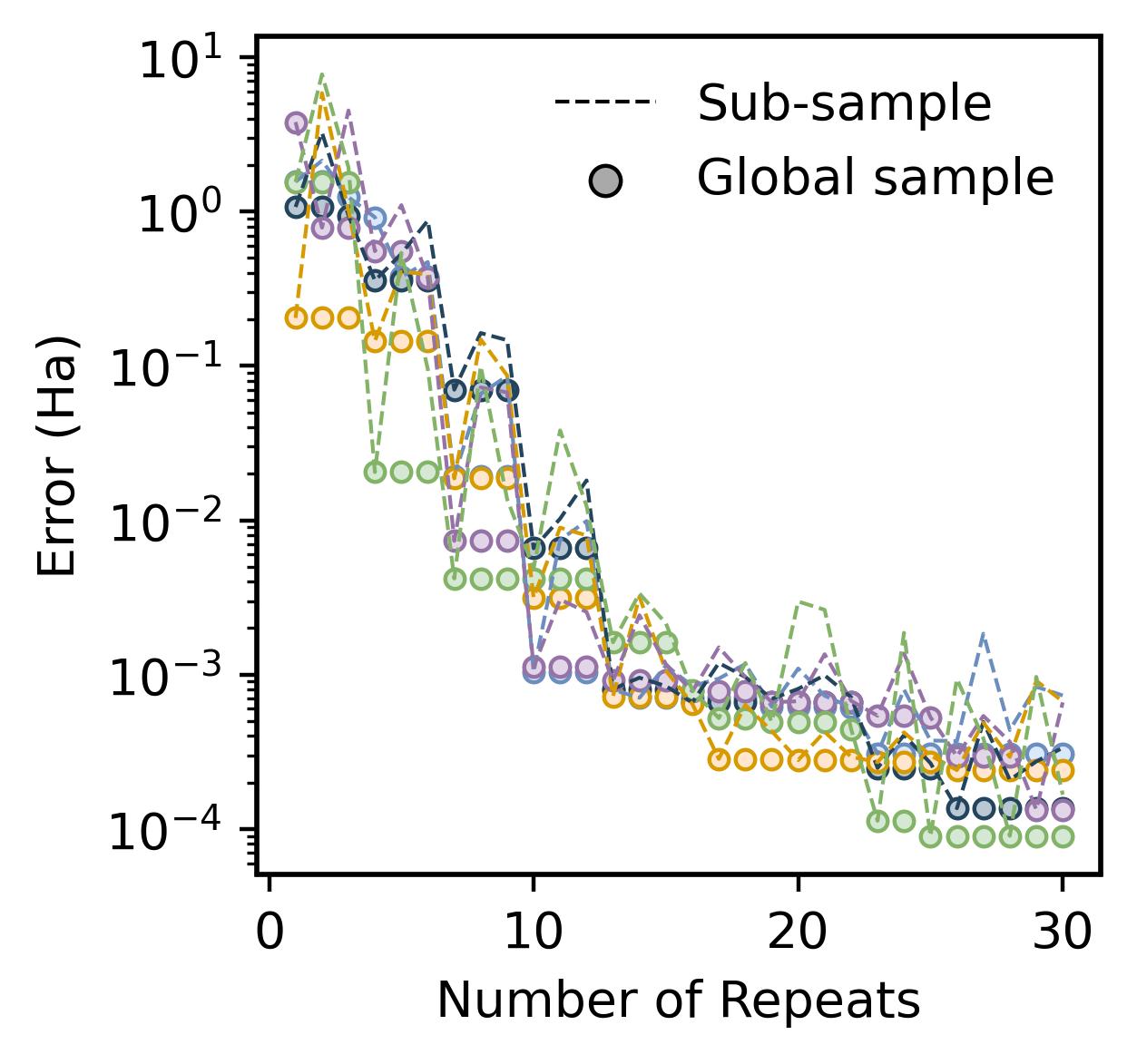} & \includegraphics[scale=0.7]{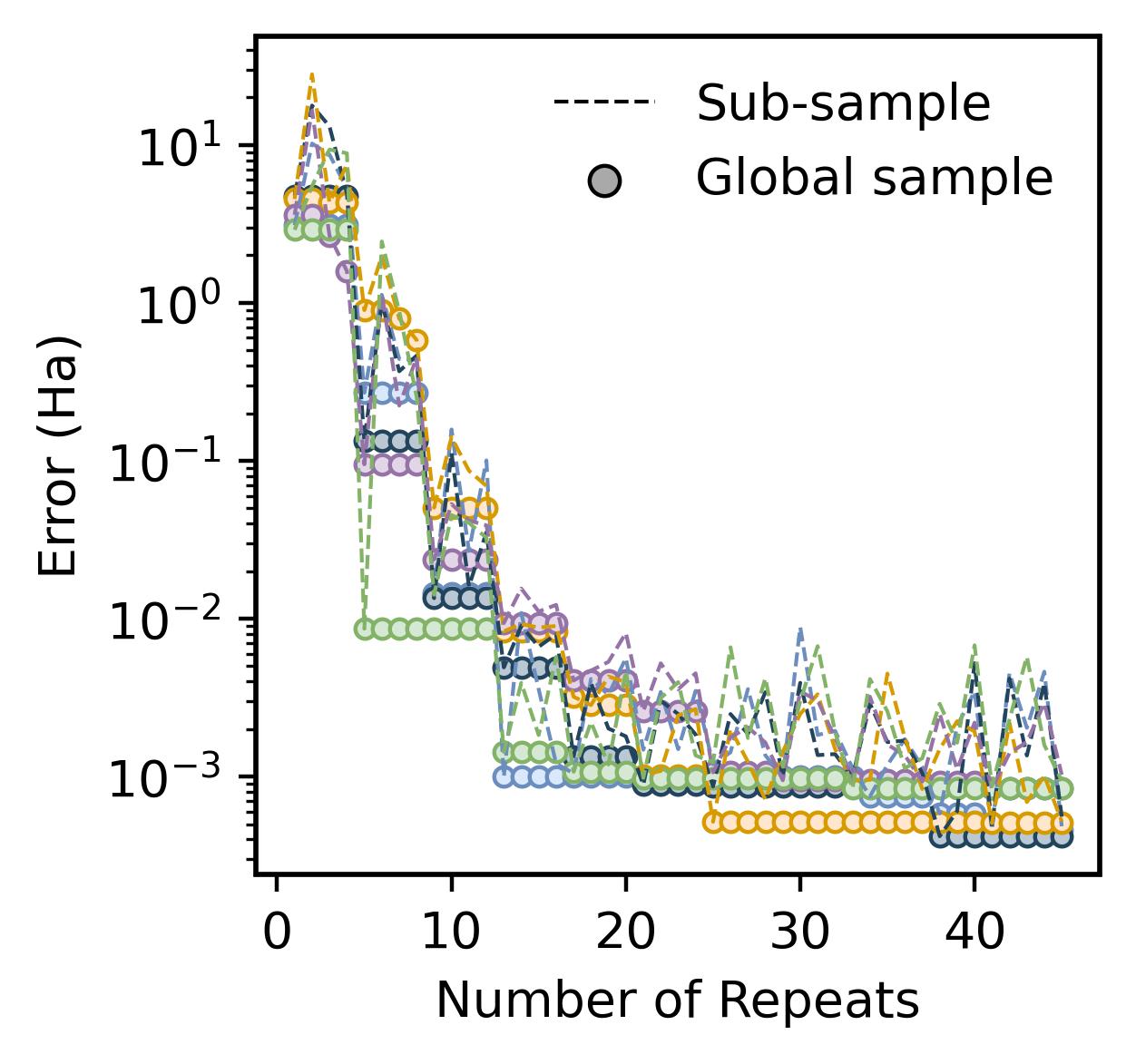}  \\
\quad \quad \quad \, (g) & \quad \quad \quad \, (h) \\

\end{tabular}
\caption{\label{fig:hwallgraph} Error in energy with respect to relCI versus number of Repeats in our workflow for $^2P_{1/2}$ (subfigures (a), (c), (d), and (f)) and $^2P_{3/2}$ (subfigures (b), (d), (f), and (h)) states of all the considered boron-like atoms using the D-Wave hardware. The dashed lines present the error at every repeat, while circles denote the error in the global sample. Different colours denote different repetitions of the experiment. Subfigures ((a),(b)), ((c),(d)), ((e),(f)) and ((g),(h)) correspond to boron-like Ca, Fe, Kr and Mo respectively.
}
\end{figure*} 
\end{document}